\documentclass[aps,prl,twocolumn,superscriptaddress,longbibliography]{revtex4-2}
\usepackage{amsmath,amssymb}
\usepackage{mathrsfs}
\usepackage[pdftex]{hyperref,graphicx}
\hypersetup{colorlinks = true, urlcolor = blue, linkcolor = blue, citecolor = blue}
\usepackage{physics}
\usepackage[dvipsnames]{xcolor}
\usepackage{bm}
\usepackage{dsfont}
\usepackage{bbold} 
\usepackage{xr}

\newcommand{\ii}{{\bf\mathrm{i}}}
\newcommand{\sfA}{\mathsf{A}}
\newcommand{\sfB}{\mathsf{B}}

\begin{document}
\title{Topological Modes in Monitored Quantum Dynamics}
\author{Haining Pan}
\affiliation{Department of Physics, Cornell University, Ithaca, New York 14853, USA}
\affiliation{Department of Physics and Astronomy, Center for Materials Theory, Rutgers University, Piscataway, New Jersey 08854 USA}
\author{Hassan Shapourian}
\affiliation{Cisco Quantum Lab, Los Angeles, California 90404, USA}
\author{Chao-Ming Jian}
\affiliation{Department of Physics, Cornell University, Ithaca, New York 14853, USA}
\begin{abstract}
Dynamical quantum systems both driven by unitary evolutions and monitored through measurements have proved to be fertile ground for exploring new dynamical quantum matters. 
While the entanglement structure and symmetry properties of monitored systems have been intensively studied, 
the role of topology in monitored dynamics is much less explored. In this work, we investigate novel topological phenomena in the monitored dynamics through the lens of free-fermion systems. 
Free-fermion monitored dynamics were previously shown to be unified with the Anderson localization problem under the Altland-Zirnbauer symmetry classification.
Guided by this unification, we identify the topological area-law-entangled phases in the former setting through the topological classification of disordered insulators and superconductors in the latter. As examples, we focus on 1+1D free-fermion monitored dynamics in two symmetry classes, DIII and A. We construct quantum circuit models to study different topological area-law phases and their domain walls in the respective symmetry classes. We find that the domain wall between topologically distinct area-law phases hosts dynamical topological modes whose entanglement is protected from being quenched by the measurements in the monitored dynamics. We demonstrate how to manipulate these topological modes by programming the domain-wall dynamics. In particular, for topological modes in class DIII, which behave as unmeasured Majorana modes, we devise a protocol to braid them and study the entanglement generated in the braiding process. 
\end{abstract}
\maketitle
\textit{Introduction.}--- Monitored dynamical quantum systems evolve through both unitary evolution and measurements that ``monitor" the systems' wave function.
The ground-breaking discovery of the measurement-induced phase transitions~\cite{skinner2019measurementinduced,li2018quantum,li2019measurementdriven,chan2019unitaryprojective,potter2022entanglement,fisher2023random} in monitored dynamics has sparked extensive research recently to explore the rich landscape of novel dynamical quantum matter in such systems. 
Significant progress has been made in understanding the changes of entanglement structure across the measurement-induced phase transitions (see, for examples, Refs.~\cite{skinner2019measurementinduced,li2018quantum,li2019measurementdriven,chan2019unitaryprojective,vasseur2019entanglement,gullans2020dynamical,gullans2020scalable,zabalo2020critical,choi2020quantum,li2021conformal,jian2020measurementinduced,bao2020theory,lang2020entanglement,turkeshi2020measurementinduced,ippoliti2021entanglement,nahum2021measurement,zabalo2022operator,li2024statistical} and reviews Refs.~\cite{potter2022entanglement,fisher2023random}) and the effect of global symmetries in enriching the phase diagram of monitored dynamical systems~\cite{sang2021measurementprotected,han2022measurementinduced,lavasani2021measurementinduced,agrawal2022entanglement,barratt2022field,jian2022criticality,jian2023measurementinduced,fava2023nonlinear,poboiko2023theory,fava2024tractable,majidy2023critical,chakraborty2023charge}. In contrast, topological properties of monitored dynamics have only been previously investigated in limited examples~\cite{lavasani2021topological,lavasani2021measurementinduced,kells2023topological,behrends2024surface}. 
This work investigates novel topological phenomena in the monitored dynamics through the lens of free-fermion systems.

In free-fermion monitored dynamics, both unitary evolutions and the measurements keep the fermionic states non-interacting. That is, a Slater-determinant state (or its charge-non-conserved counterpart) always evolves into another Slater-determinant state in every quantum trajectory (labeled by the different measurement outcomes). Despite being ``free", 
the intrinsic randomness in the measurement outcomes can still lead to rich universal behavior in the entanglement structure,
as shown in previous works~\cite{nahum2020entanglement,cao2019entanglement,kells2023topological,alberton2021entanglement,buchhold2021effective,jian2022criticality,jian2023measurementinduced,fava2023nonlinear,poboiko2023theory,poboiko2024measurementinduced,chahine2024entanglement,fava2024tractable,fidkowski2021howa,merritt2023entanglement,chen2020emergent,tang2021quantum,guo2024field,MirlinInteracting2024,behrends2024surface}. 
This work is motivated by a unification of the free-fermion monitored dynamics in $d$ spatial dimensions with the Anderson localization problem in $d+1$ spatial dimensions under the Altland-Zirnbauer (AZ) symmetry classification~\cite{altland1997nonstandard,jian2022criticality}. This unification was also recently explored in the context of quantum error correction \cite{behrends2024surface}.
An implication of this unification is a correspondence between
the phases in free-fermion monitored dynamics with an area-law entanglement entropy (EE) scaling and disordered localized phases in the Anderson localization problem of the same symmetry class. The latter is well-known to further subdivide into topological insulators and superconductors within each symmetry class
(see review Refs.~\cite{ludwig2016topological,qi2011topological}). The domain walls (DWs) between distinct topological localized phases host topologically protected modes. 
Through this correspondence, area-law phases in free-fermion dynamics should follow the same topological classification, and the DWs between them should exhibit dynamical topological domain-wall modes (DTDMs).

In this paper, 
we demonstrate the topology and the DTDMs of 1D area-law phases of free-fermion monitored dynamics in AZ symmetry classes DIII and A as examples, corresponding to 2D disordered localized phases classified by $\mathbb{Z}_2$ and $\mathbb{Z}$, respectively~\cite{ludwig2016topological,qi2011topological}.
We first construct quantum circuit models to realize the topologically distinct area-law phases, expected from the correspondence, within each symmetry class. Then, we design circuit models with DWs to investigate the entanglement structure and the topological classification of the DTDMs via numerical simulations. 
We show that despite frequent measurements in the surroundings of the DWs, the entanglement carried by the DTDMs remains protected from being quenched by the (symmetry-allowed) measurements. We demonstrate how to manipulate the DTDMs by programming DWs in the circuits. 
Specifically, for class-DIII monitored dynamics, where each DTDM is effectively an {\it unmeasured} Majorana mode, we design a protocol to braid these DTDMs and study the entanglement generated in the braiding process. 

\begin{figure}[htbp]
    \centering
    \includegraphics[width=3.4in]{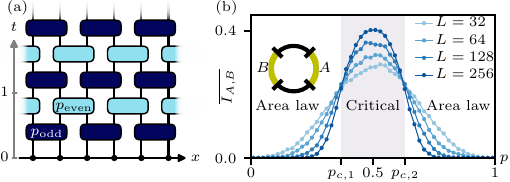}
    \caption{
        (a) Spacetime geometry of a symmetry-class-DIII monitored circuit on a Majorana chain of $L$ sites (black dots). This circuit has a staggered pattern with different measurement probabilities $p_{\rm odd/even}$ in different two-site gates.       
        (b) Average steady-state mutual information $\overline{I_{A,B}}$ as a function of $p\equiv p_{\text{odd}}=1-p_{\text{even}}$
        in the class-DIII monitored circuit shows two distinct area-law phases, separated by a critical phase. 
        Inset shows the configuration of two antipodal regions, $A$ and $B$, on a ring geometry to compute $\overline{I_{A,B}}$.
        }
    \label{fig:DIII}
\end{figure}
\textit{Monitored dynamics in a free Majorana chain}---
We start with the monitored dynamics of a free Majorana chain without any symmetry constraint (except fermion parity). This system belongs to the AZ symmetry class DIII, which can be understood from both the transfer matrix for single-particle evolution and density matrix evolution in a doubled Hilbert space
~\cite{jian2022criticality} [see Sec.~\ref{sec:correspondence} in Supplemental Material (SM) for a review]. The dynamical phases where the system evolves into states with area-law EE scaling correspond to 2D disordered class-DIII superconductors, whose topologies are $\mathbb{Z}_2$-classified. Below, we construct monitored quantum circuits to realize the two classes of area-law dynamical phases and show that each DW between them hosts a localized DTDM, equivalent to an effective {\it unmeasured} Majorana mode. We design protocols to braid these DTDMs and study the entanglement generation during the braiding process.

We consider a class-DIII free-fermion monitored dynamics driven by the monitored circuit shown in Fig.~\ref{fig:DIII}(a) acting on a chain with $L$ Majorana modes $\hat{\gamma}_{i=1,2,...,L}$ and the periodic boundary condition. For each pair of adjacent Majorana modes $(\hat{\gamma}_i,\hat{\gamma}_{i+1})$, we apply either a two-site random unitary gate, $\exp{\theta \hat{\gamma}_i\hat{\gamma}_{i+1}}$ with a random $\theta\in[0,2\pi)$, with a probability $1-p$, or a two-site projective measurement of the local fermion parity $\ii \hat{\gamma}_i\hat{\gamma}_{i+1}$ with a probability of $p$. For each measurement, the system evolves under the Kraus operators $\frac{1\pm\ii \hat{\gamma}_i\hat{\gamma}_{i+1} }{2}$ (which collapse the wave function at the measured sites) depending on the measurement outcome. Both the unitary gates and measurements evolve free-fermion states to free-fermion states. 

A staggered pattern of $p$ in the quantum circuit allows access to different dynamical phases:
In the first (second) half of each time step, the probability of measuring the parity of Majorana pairs on odd (even) links, $(\hat{\gamma}_{2i-1},\hat{\gamma}_{2i})$ [$(\hat{\gamma}_{2i},\hat{\gamma}_{2i+1})$], is denoted as $p_{\text{odd}}$ ($p_{\text{even}}$) [dark (light) blue gates in Fig.~\ref{fig:DIII}(a)].
We set $p_{\text{odd}}+p_{\text{even}}=1$ to simplify the parametrization.
When $p_{\text{odd}}$ ($p_{\text{even}}$) is large, the actions on odd (even) links are predominantly measurements, while those on the even (odd) links are predominantly two-site random unitary gates. We will determine the phase diagram of this monitored Majorana chain parametrized by $p_{\text{odd}}$ (abbreviated as $p$ hereafter).

To distinguish different dynamical phases, we numerically calculate the average steady-state mutual information (MI) $\overline{I_{A,B}}$ between two antipodal regions with $L_A=L_B=\frac{L}{4}$ as shown in Fig.~\ref{fig:DIII}(b) (see Secs.~\ref{sec:Gaussian} and \ref{sec:DIII} in SM for details).
The phase diagram in Fig.~\ref{fig:DIII}(b) illustrates two phase transitions at $p_{c,1}\approx 0.4$ and $p_{c,2}\approx 0.6$ indicated by crossings of $\overline{I_{A,B}}$ for different system sizes $L$.
For $p \in [0,p_{c,1})$ and $p \in (p_{c,2},1]$, $\overline{I_{A,B}}$ vanishes as $L$ grows, signifying the dynamical phases with the area-law scaling of the steady-state EE. For $p \in [p_{c,1}, p_{c,2}]$, the system is in a critical phase (where the EE scales logarithmically). These area-law and critical phases of the monitored Majorana chain have been previously identified by Refs.~\cite{nahum2020entanglement,jian2022criticality,jian2023measurementinduced,fava2023nonlinear,behrends2024surface} in similar models. 
We use the phase diagram in Fig.~\ref{fig:DIII}(b) to guide our subsequent investigation of the DW between the topologically distinct area-law phases at $p<p_{c,1}$ and $p>p_{c,2}$.

The physical intuition for the two topologically distinct area-law phases is as follows: In the long-time limit, the two types of dynamical phases are dominated by the measurements that dimerize the Majorana chain on the even (odd) links. Their spatial DW must host a protected DTDM, equivalent to a Majorana mode that cannot be measured by itself. 
In a complementary picture, the dynamics of the two area-law phases in the 1+1D spacetime correspond to the quantum states of the two classes of disordered topological superconductors in 2-spatial dimensions in class DIII~\cite{jian2022criticality}. Therefore, the spacetime DW between the two area-law phases must exhibit nontrivial dynamics of the DTDMs. Below, we confirm and reify these intuitions by investigating concrete circuit models with a DW between the two area-law phases. 

\begin{figure}[htbp]
    \centering
    \includegraphics[width=3.4in]{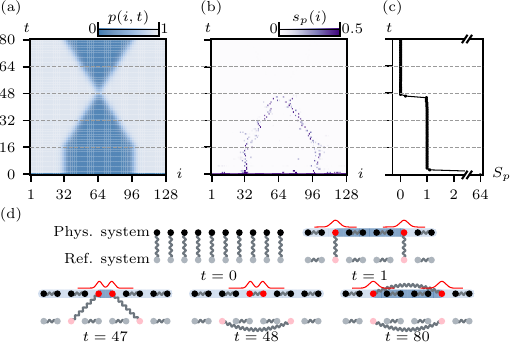}
    \caption{
        (a) Spacetime configuration $p(i,t)$ in a class-DIII monitored circuit acting on a 128-site Majorana chain. 
        (b-c) Entanglement contour and total entanglement entropy (in binary logarithm) of the physical chain in a \textit{typical} quantum trajectory. The peaks in the entanglement contour indicate the spacetime position of the dynamical topological domain-wall modes.
        (d) Schematics for the ``ideal" entanglement structure of physical and reference systems at different times $t$. Each dot presents a Majorana mode, with the DTDMs colored red. The colored strip behind physical sites represents the configuration of $p(i,t)$ shown in (a). Each pair of modes connected by a wavy line is maximally entangled.  
        } 
    \label{fig:DIII_DW}
\end{figure}
\textit{Domain-wall dynamics}---In exploring the phase diagram Fig.~\ref{fig:DIII}, we keep the $p$ homogeneous in spacetime. To study DWs and DTDMs, we generalize $p$ to a smoothly-varying spacetime-dependent probability $p(i,t)$ for measuring odd links, while locally maintaining $p_{\rm even} +p_{\rm odd}=1$. By programming $p(i,t)$, we can create spacetime domains (with $p<p_{c,1}$ and $p>p_{c,2}$) dominated by one of the two area-law dynamical phases with DWs in between. As a concrete example, we consider a 128-site Majorana chain with a spacetime configuration $p(i,t)$ shown in Fig.~\ref{fig:DIII_DW}(a). This configuration starts with a uniform $p<p_{c,1}$ at $t=0$. A pair of DWs are introduced at $i=\frac{L}{4}$ and $i=\frac{3L}{4}$ at time $t=1$, with two DTDMs expected to emerge.
For later times, we program $p(i,t)$ so that the two DWs cross and continue propagating until $t=80$.
This dynamics is robust against changing the steepness of the DW profile as the smoothness only affects the localization length of the DTDMs.

To examine the entanglement evolution in the physical Majorana chain, we introduce a reference Majorana chain of the same length [gray dots in Fig.~\ref{fig:DIII_DW}(d)] that is mode-by-mode maximally entangled with the physical chain at $t=0$. The EE $S_p$ between the physical and reference chains evolves [Fig.~\ref{fig:DIII_DW}(c)] as the physical chain undergoes the monitored circuit evolution. Since we are considering a free-fermion system, we can use the entanglement contour (EC) $s_p(i)$ to spatially resolve the physical chain's entanglement with the reference chain~\cite{chen2014entanglement}. Note that $s_p(i)$ is \textit{not} the EE at the site $i$ and satisfies $\sum_{i} s_{p}(i)=S_{p}$ (see Sec.~\ref{sec:EE} for its definition). Figure~\ref{fig:DIII_DW}(b) shows the time evolution of $s_{p}(i)$ for a typical quantum trajectory in the monitored dynamics governed by $p(i,t)$ shown in Fig.~\ref{fig:DIII_DW}(a). 

As the physical system evolves beyond $t=1$, the total EE quickly drops from the maximal value $S_{p}=\frac{L}{2}\log2=64\log2$ to $S_{p}=\log 2$. Most of the physical chain disentangles from the reference by the measurements.
The EC $s_p(i)$ develops two peaks, each contributing $\frac{1}{2}\log 2$ to the total EE, localized around the two DWs. Each peak corresponds to a DTDM [red dot in Fig.~\ref{fig:DIII_DW} (d)], equivalent to an effective Majorana mode whose entanglement with the reference chain is protected from being quenched by the local measurements, consistent with our expectation. During $t\in[17,47]$, DWs are programmed via $p(i,t)$ to approach each other, and the DTDM follows the DWs with their EC peaks still integrated to $\frac{1}{2}\log 2$ each, as shown in Fig.~\ref{fig:DIII_DW}(b).

As the two DWs cross at $t=48$, the EC peaks carried by the DTDMs annihilate each other, manifesting a $\mathbb{Z}_2$ classification consistent with that of the DWs in 2D class-DIII disordered superconductors. 
From the circuit perspective, when DWs are close, their DTDMs can be measured in pairs, causing their disentanglement from the reference chain [as illustrated in Fig.~\ref{fig:DIII_DW}(d)]. 
As DWs separate again for $t>49$, their DTDMs reemerge and become maximally entangled with each other. Hence, the EC describing the physical system's entanglement with the reference system does not detect these DTDMs for $t>49$. The circuit with $p(i,t)$ shown in Fig.~\ref{fig:DIII_DW} demonstrates the ability to control the DTDMs by programming the DW motions via $p(i,t)$. 
\begin{figure}[htbp]
    \centering
    \includegraphics[width=3.4in]{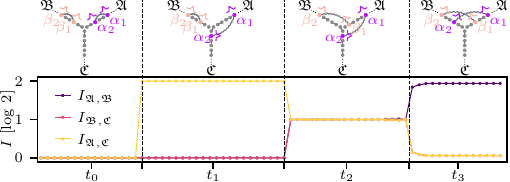}
    \caption{
    Top: Schematics for the four stages of the DTDM braiding protocol, where $\alpha_1,\alpha_2$ and $\beta_1,\beta_2$ represent the DTDMs {with finite correlation length}. We numerically simulate the protocol for a T-junction of three 64-site chains. 
    Bottom: Mutual information between the pair of chains $(\mathfrak{A}, \mathfrak{B})$, $(\mathfrak{B},\mathfrak{C})$, and $(\mathfrak{C},\mathfrak{A})$ in a \textit{typical} quantum trajectory as a function of time $t$. 
        } 
    \label{fig:braiding}
\end{figure}
\textit{Braiding.---}
Using the ability to control the DTDMs, we study a protocol to braid them in a T-junction geometry [see Fig.~\ref{fig:braiding}(a)] formed by Majorana chain $\mathfrak{A}$, $\mathfrak{B}$, and $\mathfrak{C}$. With a short-range entangled state on each chain at the beginning, the pairwise MI $I_{\mathfrak{A},\mathfrak{B}}$, $I_{\mathfrak{B},\mathfrak{C}}$, and $I_{\mathfrak{A},\mathfrak{C}}$ are all initially zero. At time $t_0$, we create a maximally-entangled pair of DTDMs $\alpha_{1,2}$ ($\beta_{1,2}$) on chain $\mathfrak{A}$ ($\mathfrak{B}$) via the dynamics shown in Fig.~\ref{fig:DIII_DW}. In the following, we devise a protocol that braids $\alpha_2$ and $\beta_1$ and study the evolution of MI $I_{\mathfrak{A},\mathfrak{B}}$, $I_{\mathfrak{B},\mathfrak{C}}$, and $I_{\mathfrak{A},\mathfrak{C}}$ among the three chains induced by the braiding in typical quantum trajectories. 

Figure~\ref{fig:braiding} outlines this braiding protocol and the evolution of $I_{\mathfrak{A},\mathfrak{B}}$, $I_{\mathfrak{B},\mathfrak{C}}$, and $I_{\mathfrak{A},\mathfrak{C}}$ in the process (See Sec.~\ref{sec:braiding} in SM for microscopic details). From time $t_0 $ to $ t_1$, we combine chain $\mathfrak{A}$ and $\mathfrak{B}$ into one single chain and program the DW configurations to move the DTDM $\alpha_2$ from chain $\mathfrak{A}$ to $\mathfrak{C}$, after which the two chains become entangled as indicated by $I_{\mathfrak{A},\mathfrak{C}}=2\log 2$ in Fig.~\ref{fig:braiding}. The DTDMs $\alpha_1$ and $\beta_{1,2}$ stand still during this time. From time $t_1 $ to $ t_2$, we use the same method to move the DTDM $\beta_1$ from chain $\mathfrak{B}$ to $\mathfrak{A}$ keeping $\alpha_{1,2}$ and $\beta_2$ fixed. From time $t_2 $ to $ t_3$, we complete the braiding by moving $\beta_1$ from chain $\mathfrak{C}$ to $\mathfrak{A}$. This braiding process entangles chains $\mathfrak{A}$ and $\mathfrak{B}$, leading to $I_{\mathfrak{A},\mathfrak{B}} = 2\log 2$ as expected. Our DTDM braiding protocol resembles the braiding protocol for Majorana zero modes using nanowire T-junctions.~\cite{alicea2011nonabelian,karzig2016universal} However, being independent of any Hamiltonian, our protocol is not limited by any energy gap in the spectrum but only the correlation length in the area-law phases. {We use $p=0.15$ to ensure the dynamics remains in area-law phase with a finite but small correlation length.}

\begin{figure}[htbp]
    \centering
    \includegraphics[width=3.4in]{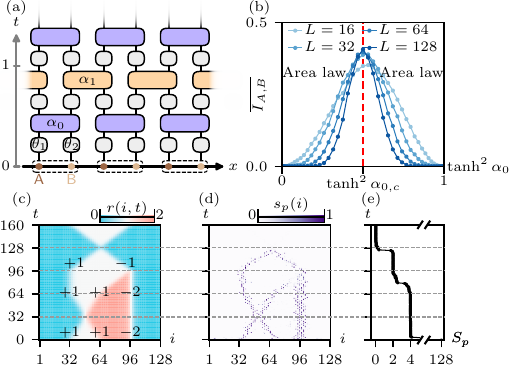}
    \caption{
        (a) Spacetime geometry of a class-A monitored circuit on a 1D complex fermion chain with each unit cell (dashed box) containing two sublattices $\sfA$ and $\sfB$. Two-site (post-selected) measurements (purple and orange gates) and onsite unitary gates (gray gates) are defined in the main text.   
        (b) Average steady-state mutual information $\overline{I_{A,B}}$ 
        as a function of $\alpha_0$ with $\tanh^2{\alpha_0}+\tanh^2\alpha_1=1$.
        (c) Spacetime configuration $r(i,t)$ for a class-A monitored circuit on a 128-site chain, encoding DWs of different integer classes as labeled.
        (d-e) Entanglement contour and total entanglement entropy (in binary logarithm) of the physical chain in a \textit{typical} quantum trajectory.
        }
    \label{fig:A}
\end{figure}
{\textit{Monitored dynamics with charge conservation and chiral symmetry---}}
Now, we study the monitored dynamics in a different AZ symmetry class, class A. The class-A area-law phases in one spatial dimension correspond to disordered insulators in the same class in two spatial dimensions~\cite{jian2022criticality}, implying the $\mathbb{Z}$ classification for both area-law phases and their DWs. Below, we study the behavior DTDMs in class-A monitored dynamics. 

A 1D monitored system in class A is constructed on complex fermions on a bipartite lattice, with sublattices $\sfA$ and $\sfB$ [dark and light brown dots in Fig.~\ref{fig:A}(a)]. We denote the fermion operators on the two sublattices in the $i$th unit cell as $c_{i,\sfA/\sfB}^\dag$ and $c_{i,\sfA/\sfB}$. Monitored dynamics in class A are defined by the dynamics that respect the U(1) charge conservation and an anti-unitary chiral symmetry ${\cal C}$, defined by $ c_{i,\sfA} \rightarrow -c_{i,\sfA}^\dagger,  c_{i,\sfB} \rightarrow c_{i,\sfB}^\dagger, \ii \rightarrow -\ii$,
in each quantum trajectory (see Sec.~\ref{sec:A} in SM for the identification of symmetry class A). In other words, the U(1) and ${\cal C}$ actions must commute with all unitary gates and Kraus operators capturing the measurements. 

To preserve U(1) and ${\cal C}$, the unitary gates must act only within each sublattice. Here, we choose the simplest case with onsite unitary gates, $\exp{\ii\theta_1\left( c_{i,\sfA}^\dagger c_{i,\sfA} -\frac{1}{2}\right)}$ and $\exp{\ii\theta_2\left( c_{j,\sfB}^\dagger c_{j,\sfB} -\frac{1}{2}\right)}$, with $\theta_{1,2}\in \mathbb{R}$ (see Sec.~\ref{sec:longrange} in SM for longer-range unitary gates). These symmetries require the Kraus operators capturing the non-unitary evolution induced by (generalized) measurements~\cite{nielsen2012quantum} to take the form $\exp{\pm\alpha( c_{i,\sfA}^\dagger c_{j,\sfB} + \text{h.c.})}$. $\alpha\in \mathbb{R}$ parametrizes the measurement strength. Such Kraus operators can be implemented as the weak measurements of the respective occupancy on the fermion modes $\frac{1}{\sqrt{2}}(c_{i,\sfA}\pm c_{j,\sfB})$ followed by post-selection of the quantum trajectories (see Sec.~\ref{sec:A} in SM for this post-selected weak measurement). The $\alpha\rightarrow \infty$ limit corresponds to the projective measurements of these particle numbers followed by post-selecting the two trajectories with a total occupancy 1 on the two associated modes.

To understand the phases of the class-A monitored dynamics, we study the circuit as shown in Fig.~\ref{fig:A}(a), where each two-site gate implements the post-selected weak measurement through the Kraus operators $\exp{\pm\alpha( c_{i,\sfA}^\dagger c_{j,\sfB} + \text{h.c.})}$ with the probability of the sign $\pm$ following the (post-selected) Born rule. To access different area-law phases, we consider different measurement strengths $\alpha_{0,1}$ for the two types of measurements (purple and orange gates) in Fig.~\ref{fig:A}(a) corresponding to $i=j$ and $i=j+1$, with $\tanh^2{\alpha_0}+\tanh^2\alpha_1=1$. Hence, when the pairs of fermion modes ($c_{i,\sfA}$, $c_{i,\sfB}$) are strongly (weakly) measured, the pairs ({$c_{i,\sfA}$, $c_{i-,\sfB}$}) are weakly (strongly) measured. Each gray gate is an onsite unitary gate defined above with independent random phases $\theta_{1,2}$. 

In Fig.~\ref{fig:A}(b), we numerically calculate the average steady-state MI $\overline{I_{A,B}}$ for two antipodal intervals $A$ and $B$ [in the same geometry as in Fig.~\ref{fig:DIII}(b)] as a function of $\alpha_0$ to map out the phase diagram. This phase diagram contains two area-law phases (separated by $\alpha_0=\alpha_{0,c}$) where the chain develops a steady-state dimerization pattern between the two sublattices within each unit cell or spanning neighboring unit cells. 
In general, there should be an area-law dynamical phase associated with the range-$R$ dimers between ($c_{i,\sfA}$, $c_{i+R,\sfB}$) for every $R\in \mathbb{Z}$, matching the $\mathbb{Z}$ classification of the corresponding 2D class-A disordered insulators~\cite{jian2022criticality,ludwig2016topological,qi2011topological}. The DW between two area-law dynamical phases with different dimerization ranges $R$ (on the left) and $R'$ (on the right) is expected to carry $\abs{R-R'}$ DTDMs, each being an effective complex-fermion mode of the same sublattice type ($\sfA$ or $\sfB$ depending on the sign of $R-R'$). The entanglement on the DTDMs is protected from being quenched by local measurements because the chiral symmetry ${\cal C}$ required by symmetry class A only allows modes from the sublattices $\sfA$ and $\sfB$ to be measured in pairs. Therefore, each DW should be classified by $\Delta R = R'-R\in \mathbb{Z}$.

To numerically study the entanglement carried by the DTDMs, we generalize the two-site measurements in Fig.~\ref{fig:A}(a) to measurements with their ranges controlled by a smoothly-varying function $r(i,t)$ in spacetime.
For the unit cell $i$ at time $t$, we measure (with post-selection) the fermion pairs ($c_{i,\sfA}, c_{i+ \lfloor r\rfloor,\sfB}$) and ($c_{i,\sfB}, c_{i+ \lfloor r\rfloor+1,\sfA}$) with respective measurement strengths $\tanh^2\alpha = 0.04, 0.96$, with a probability $\lfloor r\rfloor+1-r$.
Otherwise, we measure the pairs ($c_{i,\sfA}, c_{i+ \lfloor r\rfloor+1,\sfB}$) and ($c_{i,\sfB}, c_{i+ \lfloor r\rfloor+2,\sfA}$) with the same strengths.
Intuitively, the dominating measurement at spacetime coordinates $(i,t)$ favors the dimers on the pair ($c_{i,\sfB}, c_{i+ R,\sfA}$) with the range $R$ the closest integer to $r(i,t)+1$. 
With this generalization (and the unitary gates unchanged), we can study the class-A monitored circuit with various DWs incorporated through the $r(i,t)$ configuration shown in Fig.~\ref{fig:A}(a) on a 128-site chain. Similar to class DIII, we introduce a reference chain, maximally entangled with the physical chain at $t=0$. We calculate the EC $s_p(i)$ of the physical chain's entanglement with the reference, and the total EE $S_p$ [Figs.~\ref{fig:A}(d-e)]. 

In Fig.~\ref{fig:A}(c), we mark the topological classes $\Delta R\in \mathbb{Z}$ for each DW based on the dimerization favored by the configuration $r(i,t)$. Fig.~\ref{fig:A}(d) shows that the peaks of EC always follow the DWs with each peak's EC integrated to exactly $|\Delta R|\log2$, as expected for $|\Delta R|$ complex-fermion DTDMs. DWs with the same sign of $\Delta R$ can pass through each other, e.g., at $t=32$, with their EC peaks unaffected.
However, the EC peaks of the DWs with opposite signs of $\Delta R$ (partially) annihilate each other when the DWs meet (e.g., at $t=96,128$), reflecting that the sign of $\Delta R$ decides the sublattice character of the DTDMs. 
The EC analysis also indicate that when two DWs merge, their classes $\Delta R$'s simply add. The EC peaks no longer follow the DWs beyond $t>128$ because the physical system has disentangled from the reference. 
The circuits with longer-range unitary gates studied in Sec.~\ref{sec:longrange} of SM confirm the robustness of the DTDMs as long as the symmetries U(1) and ${\cal C}$ are preserved. This robustness is comprised when ${\cal C}$ is broken (see Sec.~\ref{sec:symm_break} in SM).

\textit{Summary and discussion---}
In this work, we study the topological properties of area-law entangled phases and their DWs in free-fermion monitored dynamics, specifically showcasing examples in AZ symmetry classes DIII and A in one spatial dimension. Via a correspondence to 2D disordered topological insulators and superconductors in the respective symmetry classes, we show that the area-law phases in class-DIII (class-A) free-fermion monitored circuits on 1D chains follow a $\mathbb{Z}_2$ ($\mathbb{Z}$) classification. The DW between different area-law phases, also classified by $\mathbb{Z}_2$ and $\mathbb{Z}$ respectively, carries DTDMs with robust entanglement protected from being quenched by the measurement in the surrounding area-law dynamics. We verify the expected topological types of DTDMs and their robustness by designing and numerically simulating the class-DIII and class-A monitored circuits with DWs. For symmetry class DIII where the DTDMs are effective unmeasured Majorana modes, we devise a braiding protocol for these DTDMs by controlling the DW dynamics, {which can be experimentally demonstrated in the fermionic quantum simulation system.~\cite{zatelli2024robust}} Our study can be readily generalized to free-fermion monitored dynamics in higher dimensions in all AZ symmetries classes. It highlights an interesting direction to explore the topology within area-law entangled dynamics phases in monitored quantum systems.

\emph{Acknowledgements.}--- C.-M.J. thanks Andreas W. W. Ludwig for collaboration on related topics. 
H.P. is supported by the National Science Foundation (Platform for the Accelerated Realization, Analysis, and Discovery of Interface Materials (PARADIM)) under Cooperative Agreement No. DMR-2039380, and US-ONR grant No.~N00014-23-1-2357.
C.-M.J. is supported by the Alfred P. Sloan Foundation through a Sloan Research Fellowship.  

\let\oldaddcontentsline\addcontentsline
\renewcommand{\addcontentsline}[3]{}
\bibliography{GTN}

\begin{thebibliography}{65}%
\makeatletter
\providecommand \@ifxundefined [1]{%
 \@ifx{#1\undefined}
}%
\providecommand \@ifnum [1]{%
 \ifnum #1\expandafter \@firstoftwo
 \else \expandafter \@secondoftwo
 \fi
}%
\providecommand \@ifx [1]{%
 \ifx #1\expandafter \@firstoftwo
 \else \expandafter \@secondoftwo
 \fi
}%
\providecommand \natexlab [1]{#1}%
\providecommand \enquote  [1]{``#1''}%
\providecommand \bibnamefont  [1]{#1}%
\providecommand \bibfnamefont [1]{#1}%
\providecommand \citenamefont [1]{#1}%
\providecommand \href@noop [0]{\@secondoftwo}%
\providecommand \href [0]{\begingroup \@sanitize@url \@href}%
\providecommand \@href[1]{\@@startlink{#1}\@@href}%
\providecommand \@@href[1]{\endgroup#1\@@endlink}%
\providecommand \@sanitize@url [0]{\catcode `\\12\catcode `\$12\catcode `\&12\catcode `\#12\catcode `\^12\catcode `\_12\catcode `\%12\relax}%
\providecommand \@@startlink[1]{}%
\providecommand \@@endlink[0]{}%
\providecommand \url  [0]{\begingroup\@sanitize@url \@url }%
\providecommand \@url [1]{\endgroup\@href {#1}{\urlprefix }}%
\providecommand \urlprefix  [0]{URL }%
\providecommand \Eprint [0]{\href }%
\providecommand \doibase [0]{https://doi.org/}%
\providecommand \selectlanguage [0]{\@gobble}%
\providecommand \bibinfo  [0]{\@secondoftwo}%
\providecommand \bibfield  [0]{\@secondoftwo}%
\providecommand \translation [1]{[#1]}%
\providecommand \BibitemOpen [0]{}%
\providecommand \bibitemStop [0]{}%
\providecommand \bibitemNoStop [0]{.\EOS\space}%
\providecommand \EOS [0]{\spacefactor3000\relax}%
\providecommand \BibitemShut  [1]{\csname bibitem#1\endcsname}%
\let\auto@bib@innerbib\@empty
\bibitem [{\citenamefont {Skinner}\ \emph {et~al.}(2019)\citenamefont {Skinner}, \citenamefont {Ruhman},\ and\ \citenamefont {Nahum}}]{skinner2019measurementinduced}%
  \BibitemOpen
  \bibfield  {author} {\bibinfo {author} {\bibfnamefont {B.}~\bibnamefont {Skinner}}, \bibinfo {author} {\bibfnamefont {J.}~\bibnamefont {Ruhman}},\ and\ \bibinfo {author} {\bibfnamefont {A.}~\bibnamefont {Nahum}},\ }\bibfield  {title} {\bibinfo {title} {Measurement-{{Induced Phase Transitions}} in the {{Dynamics}} of {{Entanglement}}},\ }\href {http://arxiv.org/abs/1808.05953} {\bibfield  {journal} {\bibinfo  {journal} {Phys. Rev. X}\ }\textbf {\bibinfo {volume} {9}},\ \bibinfo {pages} {031009} (\bibinfo {year} {2019})}\BibitemShut {NoStop}%
\bibitem [{\citenamefont {Li}\ \emph {et~al.}(2018)\citenamefont {Li}, \citenamefont {Chen},\ and\ \citenamefont {Fisher}}]{li2018quantum}%
  \BibitemOpen
  \bibfield  {author} {\bibinfo {author} {\bibfnamefont {Y.}~\bibnamefont {Li}}, \bibinfo {author} {\bibfnamefont {X.}~\bibnamefont {Chen}},\ and\ \bibinfo {author} {\bibfnamefont {M.~P.~A.}\ \bibnamefont {Fisher}},\ }\bibfield  {title} {\bibinfo {title} {Quantum {{Zeno}} effect and the many-body entanglement transition},\ }\href {https://link.aps.org/doi/10.1103/PhysRevB.98.205136} {\bibfield  {journal} {\bibinfo  {journal} {Phys. Rev. B}\ }\textbf {\bibinfo {volume} {98}},\ \bibinfo {pages} {205136} (\bibinfo {year} {2018})}\BibitemShut {NoStop}%
\bibitem [{\citenamefont {Li}\ \emph {et~al.}(2019)\citenamefont {Li}, \citenamefont {Chen},\ and\ \citenamefont {Fisher}}]{li2019measurementdriven}%
  \BibitemOpen
  \bibfield  {author} {\bibinfo {author} {\bibfnamefont {Y.}~\bibnamefont {Li}}, \bibinfo {author} {\bibfnamefont {X.}~\bibnamefont {Chen}},\ and\ \bibinfo {author} {\bibfnamefont {M.~P.~A.}\ \bibnamefont {Fisher}},\ }\bibfield  {title} {\bibinfo {title} {Measurement-driven entanglement transition in hybrid quantum circuits},\ }\href {http://arxiv.org/abs/1901.08092} {\bibfield  {journal} {\bibinfo  {journal} {Phys. Rev. B}\ }\textbf {\bibinfo {volume} {100}},\ \bibinfo {pages} {134306} (\bibinfo {year} {2019})}\BibitemShut {NoStop}%
\bibitem [{\citenamefont {Chan}\ \emph {et~al.}(2019)\citenamefont {Chan}, \citenamefont {Nandkishore}, \citenamefont {Pretko},\ and\ \citenamefont {Smith}}]{chan2019unitaryprojective}%
  \BibitemOpen
  \bibfield  {author} {\bibinfo {author} {\bibfnamefont {A.}~\bibnamefont {Chan}}, \bibinfo {author} {\bibfnamefont {R.~M.}\ \bibnamefont {Nandkishore}}, \bibinfo {author} {\bibfnamefont {M.}~\bibnamefont {Pretko}},\ and\ \bibinfo {author} {\bibfnamefont {G.}~\bibnamefont {Smith}},\ }\bibfield  {title} {\bibinfo {title} {Unitary-projective entanglement dynamics},\ }\href {http://arxiv.org/abs/1808.05949} {\bibfield  {journal} {\bibinfo  {journal} {Phys. Rev. B}\ }\textbf {\bibinfo {volume} {99}},\ \bibinfo {pages} {224307} (\bibinfo {year} {2019})}\BibitemShut {NoStop}%
\bibitem [{\citenamefont {Potter}\ and\ \citenamefont {Vasseur}(2022)}]{potter2022entanglement}%
  \BibitemOpen
  \bibfield  {author} {\bibinfo {author} {\bibfnamefont {A.~C.}\ \bibnamefont {Potter}}\ and\ \bibinfo {author} {\bibfnamefont {R.}~\bibnamefont {Vasseur}},\ }\bibfield  {title} {\bibinfo {title} {Entanglement {{Dynamics}} in {{Hybrid Quantum Circuits}}},\ }in\ \href {https://doi.org/10.1007/978-3-031-03998-0_9} {\emph {\bibinfo {booktitle} {Entanglement in {{Spin Chains}}: {{From Theory}} to {{Quantum Technology Applications}}}}},\ \bibinfo {editor} {edited by\ \bibinfo {editor} {\bibfnamefont {A.}~\bibnamefont {Bayat}}, \bibinfo {editor} {\bibfnamefont {S.}~\bibnamefont {Bose}},\ and\ \bibinfo {editor} {\bibfnamefont {H.}~\bibnamefont {Johannesson}}}\ (\bibinfo  {publisher} {Springer International Publishing},\ \bibinfo {address} {Cham},\ \bibinfo {year} {2022})\ pp.\ \bibinfo {pages} {211--249}\BibitemShut {NoStop}%
\bibitem [{\citenamefont {Fisher}\ \emph {et~al.}(2023)\citenamefont {Fisher}, \citenamefont {Khemani}, \citenamefont {Nahum},\ and\ \citenamefont {Vijay}}]{fisher2023random}%
  \BibitemOpen
  \bibfield  {author} {\bibinfo {author} {\bibfnamefont {M.~P.~A.}\ \bibnamefont {Fisher}}, \bibinfo {author} {\bibfnamefont {V.}~\bibnamefont {Khemani}}, \bibinfo {author} {\bibfnamefont {A.}~\bibnamefont {Nahum}},\ and\ \bibinfo {author} {\bibfnamefont {S.}~\bibnamefont {Vijay}},\ }\bibfield  {title} {\bibinfo {title} {Random {{Quantum Circuits}}},\ }\href {https://www.annualreviews.org/content/journals/10.1146/annurev-conmatphys-031720-030658} {\bibfield  {journal} {\bibinfo  {journal} {Annual Review of Condensed Matter Physics}\ }\textbf {\bibinfo {volume} {14}},\ \bibinfo {pages} {335} (\bibinfo {year} {2023})}\BibitemShut {NoStop}%
\bibitem [{\citenamefont {Vasseur}\ \emph {et~al.}(2019)\citenamefont {Vasseur}, \citenamefont {Potter}, \citenamefont {You},\ and\ \citenamefont {Ludwig}}]{vasseur2019entanglement}%
  \BibitemOpen
  \bibfield  {author} {\bibinfo {author} {\bibfnamefont {R.}~\bibnamefont {Vasseur}}, \bibinfo {author} {\bibfnamefont {A.~C.}\ \bibnamefont {Potter}}, \bibinfo {author} {\bibfnamefont {Y.-Z.}\ \bibnamefont {You}},\ and\ \bibinfo {author} {\bibfnamefont {A.~W.~W.}\ \bibnamefont {Ludwig}},\ }\bibfield  {title} {\bibinfo {title} {Entanglement {{Transitions}} from {{Holographic Random Tensor Networks}}},\ }\href {http://arxiv.org/abs/1807.07082} {\bibfield  {journal} {\bibinfo  {journal} {Phys. Rev. B}\ }\textbf {\bibinfo {volume} {100}},\ \bibinfo {pages} {134203} (\bibinfo {year} {2019})}\BibitemShut {NoStop}%
\bibitem [{\citenamefont {Gullans}\ and\ \citenamefont {Huse}(2020{\natexlab{a}})}]{gullans2020dynamical}%
  \BibitemOpen
  \bibfield  {author} {\bibinfo {author} {\bibfnamefont {M.~J.}\ \bibnamefont {Gullans}}\ and\ \bibinfo {author} {\bibfnamefont {D.~A.}\ \bibnamefont {Huse}},\ }\bibfield  {title} {\bibinfo {title} {Dynamical {{Purification Phase Transition Induced}} by {{Quantum Measurements}}},\ }\href {https://link.aps.org/doi/10.1103/PhysRevX.10.041020} {\bibfield  {journal} {\bibinfo  {journal} {Phys. Rev. X}\ }\textbf {\bibinfo {volume} {10}},\ \bibinfo {pages} {041020} (\bibinfo {year} {2020}{\natexlab{a}})}\BibitemShut {NoStop}%
\bibitem [{\citenamefont {Gullans}\ and\ \citenamefont {Huse}(2020{\natexlab{b}})}]{gullans2020scalable}%
  \BibitemOpen
  \bibfield  {author} {\bibinfo {author} {\bibfnamefont {M.~J.}\ \bibnamefont {Gullans}}\ and\ \bibinfo {author} {\bibfnamefont {D.~A.}\ \bibnamefont {Huse}},\ }\bibfield  {title} {\bibinfo {title} {Scalable {{Probes}} of {{Measurement-Induced Criticality}}},\ }\href {https://link.aps.org/doi/10.1103/PhysRevLett.125.070606} {\bibfield  {journal} {\bibinfo  {journal} {Phys. Rev. Lett.}\ }\textbf {\bibinfo {volume} {125}},\ \bibinfo {pages} {070606} (\bibinfo {year} {2020}{\natexlab{b}})}\BibitemShut {NoStop}%
\bibitem [{\citenamefont {Zabalo}\ \emph {et~al.}(2020)\citenamefont {Zabalo}, \citenamefont {Gullans}, \citenamefont {Wilson}, \citenamefont {Gopalakrishnan}, \citenamefont {Huse},\ and\ \citenamefont {Pixley}}]{zabalo2020critical}%
  \BibitemOpen
  \bibfield  {author} {\bibinfo {author} {\bibfnamefont {A.}~\bibnamefont {Zabalo}}, \bibinfo {author} {\bibfnamefont {M.~J.}\ \bibnamefont {Gullans}}, \bibinfo {author} {\bibfnamefont {J.~H.}\ \bibnamefont {Wilson}}, \bibinfo {author} {\bibfnamefont {S.}~\bibnamefont {Gopalakrishnan}}, \bibinfo {author} {\bibfnamefont {D.~A.}\ \bibnamefont {Huse}},\ and\ \bibinfo {author} {\bibfnamefont {J.~H.}\ \bibnamefont {Pixley}},\ }\bibfield  {title} {\bibinfo {title} {Critical properties of the measurement-induced transition in random quantum circuits},\ }\href {https://link.aps.org/doi/10.1103/PhysRevB.101.060301} {\bibfield  {journal} {\bibinfo  {journal} {Phys. Rev. B}\ }\textbf {\bibinfo {volume} {101}},\ \bibinfo {pages} {060301} (\bibinfo {year} {2020})}\BibitemShut {NoStop}%
\bibitem [{\citenamefont {Choi}\ \emph {et~al.}(2020)\citenamefont {Choi}, \citenamefont {Bao}, \citenamefont {Qi},\ and\ \citenamefont {Altman}}]{choi2020quantum}%
  \BibitemOpen
  \bibfield  {author} {\bibinfo {author} {\bibfnamefont {S.}~\bibnamefont {Choi}}, \bibinfo {author} {\bibfnamefont {Y.}~\bibnamefont {Bao}}, \bibinfo {author} {\bibfnamefont {X.-L.}\ \bibnamefont {Qi}},\ and\ \bibinfo {author} {\bibfnamefont {E.}~\bibnamefont {Altman}},\ }\bibfield  {title} {\bibinfo {title} {Quantum {{Error Correction}} in {{Scrambling Dynamics}} and {{Measurement-Induced Phase Transition}}},\ }\href {https://link.aps.org/doi/10.1103/PhysRevLett.125.030505} {\bibfield  {journal} {\bibinfo  {journal} {Phys. Rev. Lett.}\ }\textbf {\bibinfo {volume} {125}},\ \bibinfo {pages} {030505} (\bibinfo {year} {2020})}\BibitemShut {NoStop}%
\bibitem [{\citenamefont {Li}\ \emph {et~al.}(2021)\citenamefont {Li}, \citenamefont {Chen}, \citenamefont {Ludwig},\ and\ \citenamefont {Fisher}}]{li2021conformal}%
  \BibitemOpen
  \bibfield  {author} {\bibinfo {author} {\bibfnamefont {Y.}~\bibnamefont {Li}}, \bibinfo {author} {\bibfnamefont {X.}~\bibnamefont {Chen}}, \bibinfo {author} {\bibfnamefont {A.~W.~W.}\ \bibnamefont {Ludwig}},\ and\ \bibinfo {author} {\bibfnamefont {M.~P.~A.}\ \bibnamefont {Fisher}},\ }\bibfield  {title} {\bibinfo {title} {Conformal invariance and quantum nonlocality in critical hybrid circuits},\ }\href {https://link.aps.org/doi/10.1103/PhysRevB.104.104305} {\bibfield  {journal} {\bibinfo  {journal} {Phys. Rev. B}\ }\textbf {\bibinfo {volume} {104}},\ \bibinfo {pages} {104305} (\bibinfo {year} {2021})}\BibitemShut {NoStop}%
\bibitem [{\citenamefont {Jian}\ \emph {et~al.}(2020)\citenamefont {Jian}, \citenamefont {You}, \citenamefont {Vasseur},\ and\ \citenamefont {Ludwig}}]{jian2020measurementinduced}%
  \BibitemOpen
  \bibfield  {author} {\bibinfo {author} {\bibfnamefont {C.-M.}\ \bibnamefont {Jian}}, \bibinfo {author} {\bibfnamefont {Y.-Z.}\ \bibnamefont {You}}, \bibinfo {author} {\bibfnamefont {R.}~\bibnamefont {Vasseur}},\ and\ \bibinfo {author} {\bibfnamefont {A.~W.~W.}\ \bibnamefont {Ludwig}},\ }\bibfield  {title} {\bibinfo {title} {Measurement-induced criticality in random quantum circuits},\ }\href {https://link.aps.org/doi/10.1103/PhysRevB.101.104302} {\bibfield  {journal} {\bibinfo  {journal} {Phys. Rev. B}\ }\textbf {\bibinfo {volume} {101}},\ \bibinfo {pages} {104302} (\bibinfo {year} {2020})}\BibitemShut {NoStop}%
\bibitem [{\citenamefont {Bao}\ \emph {et~al.}(2020)\citenamefont {Bao}, \citenamefont {Choi},\ and\ \citenamefont {Altman}}]{bao2020theory}%
  \BibitemOpen
  \bibfield  {author} {\bibinfo {author} {\bibfnamefont {Y.}~\bibnamefont {Bao}}, \bibinfo {author} {\bibfnamefont {S.}~\bibnamefont {Choi}},\ and\ \bibinfo {author} {\bibfnamefont {E.}~\bibnamefont {Altman}},\ }\bibfield  {title} {\bibinfo {title} {Theory of the phase transition in random unitary circuits with measurements},\ }\href {https://link.aps.org/doi/10.1103/PhysRevB.101.104301} {\bibfield  {journal} {\bibinfo  {journal} {Phys. Rev. B}\ }\textbf {\bibinfo {volume} {101}},\ \bibinfo {pages} {104301} (\bibinfo {year} {2020})}\BibitemShut {NoStop}%
\bibitem [{\citenamefont {Lang}\ and\ \citenamefont {B{\"u}chler}(2020)}]{lang2020entanglement}%
  \BibitemOpen
  \bibfield  {author} {\bibinfo {author} {\bibfnamefont {N.}~\bibnamefont {Lang}}\ and\ \bibinfo {author} {\bibfnamefont {H.~P.}\ \bibnamefont {B{\"u}chler}},\ }\bibfield  {title} {\bibinfo {title} {Entanglement transition in the projective transverse field {{Ising}} model},\ }\href {https://link.aps.org/doi/10.1103/PhysRevB.102.094204} {\bibfield  {journal} {\bibinfo  {journal} {Phys. Rev. B}\ }\textbf {\bibinfo {volume} {102}},\ \bibinfo {pages} {094204} (\bibinfo {year} {2020})}\BibitemShut {NoStop}%
\bibitem [{\citenamefont {Turkeshi}\ \emph {et~al.}(2020)\citenamefont {Turkeshi}, \citenamefont {Fazio},\ and\ \citenamefont {Dalmonte}}]{turkeshi2020measurementinduced}%
  \BibitemOpen
  \bibfield  {author} {\bibinfo {author} {\bibfnamefont {X.}~\bibnamefont {Turkeshi}}, \bibinfo {author} {\bibfnamefont {R.}~\bibnamefont {Fazio}},\ and\ \bibinfo {author} {\bibfnamefont {M.}~\bibnamefont {Dalmonte}},\ }\bibfield  {title} {\bibinfo {title} {Measurement-induced criticality in {$(2+1)$}-dimensional hybrid quantum circuits},\ }\href {https://link.aps.org/doi/10.1103/PhysRevB.102.014315} {\bibfield  {journal} {\bibinfo  {journal} {Phys. Rev. B}\ }\textbf {\bibinfo {volume} {102}},\ \bibinfo {pages} {014315} (\bibinfo {year} {2020})}\BibitemShut {NoStop}%
\bibitem [{\citenamefont {Ippoliti}\ \emph {et~al.}(2021)\citenamefont {Ippoliti}, \citenamefont {Gullans}, \citenamefont {Gopalakrishnan}, \citenamefont {Huse},\ and\ \citenamefont {Khemani}}]{ippoliti2021entanglement}%
  \BibitemOpen
  \bibfield  {author} {\bibinfo {author} {\bibfnamefont {M.}~\bibnamefont {Ippoliti}}, \bibinfo {author} {\bibfnamefont {M.~J.}\ \bibnamefont {Gullans}}, \bibinfo {author} {\bibfnamefont {S.}~\bibnamefont {Gopalakrishnan}}, \bibinfo {author} {\bibfnamefont {D.~A.}\ \bibnamefont {Huse}},\ and\ \bibinfo {author} {\bibfnamefont {V.}~\bibnamefont {Khemani}},\ }\bibfield  {title} {\bibinfo {title} {Entanglement {{Phase Transitions}} in {{Measurement-Only Dynamics}}},\ }\href {https://link.aps.org/doi/10.1103/PhysRevX.11.011030} {\bibfield  {journal} {\bibinfo  {journal} {Phys. Rev. X}\ }\textbf {\bibinfo {volume} {11}},\ \bibinfo {pages} {011030} (\bibinfo {year} {2021})}\BibitemShut {NoStop}%
\bibitem [{\citenamefont {Nahum}\ \emph {et~al.}(2021)\citenamefont {Nahum}, \citenamefont {Roy}, \citenamefont {Skinner},\ and\ \citenamefont {Ruhman}}]{nahum2021measurement}%
  \BibitemOpen
  \bibfield  {author} {\bibinfo {author} {\bibfnamefont {A.}~\bibnamefont {Nahum}}, \bibinfo {author} {\bibfnamefont {S.}~\bibnamefont {Roy}}, \bibinfo {author} {\bibfnamefont {B.}~\bibnamefont {Skinner}},\ and\ \bibinfo {author} {\bibfnamefont {J.}~\bibnamefont {Ruhman}},\ }\bibfield  {title} {\bibinfo {title} {Measurement and {{Entanglement Phase Transitions}} in {{All-To-All Quantum Circuits}}, on {{Quantum Trees}}, and in {{Landau-Ginsburg Theory}}},\ }\href {https://link.aps.org/doi/10.1103/PRXQuantum.2.010352} {\bibfield  {journal} {\bibinfo  {journal} {PRX Quantum}\ }\textbf {\bibinfo {volume} {2}},\ \bibinfo {pages} {010352} (\bibinfo {year} {2021})}\BibitemShut {NoStop}%
\bibitem [{\citenamefont {Zabalo}\ \emph {et~al.}(2022)\citenamefont {Zabalo}, \citenamefont {Gullans}, \citenamefont {Wilson}, \citenamefont {Vasseur}, \citenamefont {Ludwig}, \citenamefont {Gopalakrishnan}, \citenamefont {Huse},\ and\ \citenamefont {Pixley}}]{zabalo2022operator}%
  \BibitemOpen
  \bibfield  {author} {\bibinfo {author} {\bibfnamefont {A.}~\bibnamefont {Zabalo}}, \bibinfo {author} {\bibfnamefont {M.~J.}\ \bibnamefont {Gullans}}, \bibinfo {author} {\bibfnamefont {J.~H.}\ \bibnamefont {Wilson}}, \bibinfo {author} {\bibfnamefont {R.}~\bibnamefont {Vasseur}}, \bibinfo {author} {\bibfnamefont {A.~W.~W.}\ \bibnamefont {Ludwig}}, \bibinfo {author} {\bibfnamefont {S.}~\bibnamefont {Gopalakrishnan}}, \bibinfo {author} {\bibfnamefont {D.~A.}\ \bibnamefont {Huse}},\ and\ \bibinfo {author} {\bibfnamefont {J.~H.}\ \bibnamefont {Pixley}},\ }\bibfield  {title} {\bibinfo {title} {Operator {{Scaling Dimensions}} and {{Multifractality}} at {{Measurement-Induced Transitions}}},\ }\href {https://link.aps.org/doi/10.1103/PhysRevLett.128.050602} {\bibfield  {journal} {\bibinfo  {journal} {Phys. Rev. Lett.}\ }\textbf {\bibinfo {volume} {128}},\ \bibinfo {pages} {050602} (\bibinfo {year} {2022})}\BibitemShut {NoStop}%
\bibitem [{\citenamefont {Li}\ \emph {et~al.}(2024)\citenamefont {Li}, \citenamefont {Vasseur}, \citenamefont {Fisher},\ and\ \citenamefont {Ludwig}}]{li2024statistical}%
  \BibitemOpen
  \bibfield  {author} {\bibinfo {author} {\bibfnamefont {Y.}~\bibnamefont {Li}}, \bibinfo {author} {\bibfnamefont {R.}~\bibnamefont {Vasseur}}, \bibinfo {author} {\bibfnamefont {M.~P.~A.}\ \bibnamefont {Fisher}},\ and\ \bibinfo {author} {\bibfnamefont {A.~W.~W.}\ \bibnamefont {Ludwig}},\ }\bibfield  {title} {\bibinfo {title} {Statistical {{Mechanics Model}} for {{Clifford Random Tensor Networks}} and {{Monitored Quantum Circuits}}},\ }\href {http://arxiv.org/abs/2110.02988} {\bibfield  {journal} {\bibinfo  {journal} {Phys. Rev. B}\ }\textbf {\bibinfo {volume} {109}},\ \bibinfo {pages} {174307} (\bibinfo {year} {2024})}\BibitemShut {NoStop}%
\bibitem [{\citenamefont {Sang}\ and\ \citenamefont {Hsieh}(2021)}]{sang2021measurementprotected}%
  \BibitemOpen
  \bibfield  {author} {\bibinfo {author} {\bibfnamefont {S.}~\bibnamefont {Sang}}\ and\ \bibinfo {author} {\bibfnamefont {T.~H.}\ \bibnamefont {Hsieh}},\ }\bibfield  {title} {\bibinfo {title} {Measurement-protected quantum phases},\ }\href {https://link.aps.org/doi/10.1103/PhysRevResearch.3.023200} {\bibfield  {journal} {\bibinfo  {journal} {Phys. Rev. Res.}\ }\textbf {\bibinfo {volume} {3}},\ \bibinfo {pages} {023200} (\bibinfo {year} {2021})}\BibitemShut {NoStop}%
\bibitem [{\citenamefont {Han}\ and\ \citenamefont {Chen}(2022)}]{han2022measurementinduced}%
  \BibitemOpen
  \bibfield  {author} {\bibinfo {author} {\bibfnamefont {Y.}~\bibnamefont {Han}}\ and\ \bibinfo {author} {\bibfnamefont {X.}~\bibnamefont {Chen}},\ }\bibfield  {title} {\bibinfo {title} {Measurement-induced criticality in {${\mathbb{Z}}_{2}$}-symmetric quantum automaton circuits},\ }\href {https://link.aps.org/doi/10.1103/PhysRevB.105.064306} {\bibfield  {journal} {\bibinfo  {journal} {Phys. Rev. B}\ }\textbf {\bibinfo {volume} {105}},\ \bibinfo {pages} {064306} (\bibinfo {year} {2022})}\BibitemShut {NoStop}%
\bibitem [{\citenamefont {Lavasani}\ \emph {et~al.}(2021{\natexlab{a}})\citenamefont {Lavasani}, \citenamefont {Alavirad},\ and\ \citenamefont {Barkeshli}}]{lavasani2021measurementinduced}%
  \BibitemOpen
  \bibfield  {author} {\bibinfo {author} {\bibfnamefont {A.}~\bibnamefont {Lavasani}}, \bibinfo {author} {\bibfnamefont {Y.}~\bibnamefont {Alavirad}},\ and\ \bibinfo {author} {\bibfnamefont {M.}~\bibnamefont {Barkeshli}},\ }\bibfield  {title} {\bibinfo {title} {Measurement-induced topological entanglement transitions in symmetric random quantum circuits},\ }\href {https://www.nature.com/articles/s41567-020-01112-z} {\bibfield  {journal} {\bibinfo  {journal} {Nat. Phys.}\ }\textbf {\bibinfo {volume} {17}},\ \bibinfo {pages} {342} (\bibinfo {year} {2021}{\natexlab{a}})}\BibitemShut {NoStop}%
\bibitem [{\citenamefont {Agrawal}\ \emph {et~al.}(2022)\citenamefont {Agrawal}, \citenamefont {Zabalo}, \citenamefont {Chen}, \citenamefont {Wilson}, \citenamefont {Potter}, \citenamefont {Pixley}, \citenamefont {Gopalakrishnan},\ and\ \citenamefont {Vasseur}}]{agrawal2022entanglement}%
  \BibitemOpen
  \bibfield  {author} {\bibinfo {author} {\bibfnamefont {U.}~\bibnamefont {Agrawal}}, \bibinfo {author} {\bibfnamefont {A.}~\bibnamefont {Zabalo}}, \bibinfo {author} {\bibfnamefont {K.}~\bibnamefont {Chen}}, \bibinfo {author} {\bibfnamefont {J.~H.}\ \bibnamefont {Wilson}}, \bibinfo {author} {\bibfnamefont {A.~C.}\ \bibnamefont {Potter}}, \bibinfo {author} {\bibfnamefont {J.~H.}\ \bibnamefont {Pixley}}, \bibinfo {author} {\bibfnamefont {S.}~\bibnamefont {Gopalakrishnan}},\ and\ \bibinfo {author} {\bibfnamefont {R.}~\bibnamefont {Vasseur}},\ }\bibfield  {title} {\bibinfo {title} {Entanglement and charge-sharpening transitions in {{U}}(1) symmetric monitored quantum circuits},\ }\href {http://arxiv.org/abs/2107.10279} {\bibfield  {journal} {\bibinfo  {journal} {Phys. Rev. X}\ }\textbf {\bibinfo {volume} {12}},\ \bibinfo {pages} {041002} (\bibinfo {year} {2022})}\BibitemShut {NoStop}%
\bibitem [{\citenamefont {Barratt}\ \emph {et~al.}(2022)\citenamefont {Barratt}, \citenamefont {Agrawal}, \citenamefont {Gopalakrishnan}, \citenamefont {Huse}, \citenamefont {Vasseur},\ and\ \citenamefont {Potter}}]{barratt2022field}%
  \BibitemOpen
  \bibfield  {author} {\bibinfo {author} {\bibfnamefont {F.}~\bibnamefont {Barratt}}, \bibinfo {author} {\bibfnamefont {U.}~\bibnamefont {Agrawal}}, \bibinfo {author} {\bibfnamefont {S.}~\bibnamefont {Gopalakrishnan}}, \bibinfo {author} {\bibfnamefont {D.~A.}\ \bibnamefont {Huse}}, \bibinfo {author} {\bibfnamefont {R.}~\bibnamefont {Vasseur}},\ and\ \bibinfo {author} {\bibfnamefont {A.~C.}\ \bibnamefont {Potter}},\ }\bibfield  {title} {\bibinfo {title} {Field theory of charge sharpening in symmetric monitored quantum circuits},\ }\href {http://arxiv.org/abs/2111.09336} {\bibfield  {journal} {\bibinfo  {journal} {Phys. Rev. Lett.}\ }\textbf {\bibinfo {volume} {129}},\ \bibinfo {pages} {120604} (\bibinfo {year} {2022})}\BibitemShut {NoStop}%
\bibitem [{\citenamefont {Jian}\ \emph {et~al.}(2022)\citenamefont {Jian}, \citenamefont {Bauer}, \citenamefont {Keselman},\ and\ \citenamefont {Ludwig}}]{jian2022criticality}%
  \BibitemOpen
  \bibfield  {author} {\bibinfo {author} {\bibfnamefont {C.-M.}\ \bibnamefont {Jian}}, \bibinfo {author} {\bibfnamefont {B.}~\bibnamefont {Bauer}}, \bibinfo {author} {\bibfnamefont {A.}~\bibnamefont {Keselman}},\ and\ \bibinfo {author} {\bibfnamefont {A.~W.~W.}\ \bibnamefont {Ludwig}},\ }\bibfield  {title} {\bibinfo {title} {Criticality and entanglement in nonunitary quantum circuits and tensor networks of noninteracting fermions},\ }\href {https://link.aps.org/doi/10.1103/PhysRevB.106.134206} {\bibfield  {journal} {\bibinfo  {journal} {Phys. Rev. B}\ }\textbf {\bibinfo {volume} {106}},\ \bibinfo {pages} {134206} (\bibinfo {year} {2022})}\BibitemShut {NoStop}%
\bibitem [{\citenamefont {Jian}\ \emph {et~al.}(2023)\citenamefont {Jian}, \citenamefont {Shapourian}, \citenamefont {Bauer},\ and\ \citenamefont {Ludwig}}]{jian2023measurementinduced}%
  \BibitemOpen
  \bibfield  {author} {\bibinfo {author} {\bibfnamefont {C.-M.}\ \bibnamefont {Jian}}, \bibinfo {author} {\bibfnamefont {H.}~\bibnamefont {Shapourian}}, \bibinfo {author} {\bibfnamefont {B.}~\bibnamefont {Bauer}},\ and\ \bibinfo {author} {\bibfnamefont {A.~W.~W.}\ \bibnamefont {Ludwig}},\ }\bibfield  {title} {\bibinfo {title} {Measurement-induced entanglement transitions in quantum circuits of non-interacting fermions: {{Born-rule}} versus forced measurements},\ }\href {http://arxiv.org/abs/2302.09094} {\bibfield  {journal} {\bibinfo  {journal} {arXiv:2302.09094}\ } (\bibinfo {year} {2023})}\BibitemShut {NoStop}%
\bibitem [{\citenamefont {Fava}\ \emph {et~al.}(2023)\citenamefont {Fava}, \citenamefont {Piroli}, \citenamefont {Swann}, \citenamefont {Bernard},\ and\ \citenamefont {Nahum}}]{fava2023nonlinear}%
  \BibitemOpen
  \bibfield  {author} {\bibinfo {author} {\bibfnamefont {M.}~\bibnamefont {Fava}}, \bibinfo {author} {\bibfnamefont {L.}~\bibnamefont {Piroli}}, \bibinfo {author} {\bibfnamefont {T.}~\bibnamefont {Swann}}, \bibinfo {author} {\bibfnamefont {D.}~\bibnamefont {Bernard}},\ and\ \bibinfo {author} {\bibfnamefont {A.}~\bibnamefont {Nahum}},\ }\bibfield  {title} {\bibinfo {title} {Nonlinear {{Sigma Models}} for {{Monitored Dynamics}} of {{Free Fermions}}},\ }\href {https://link.aps.org/doi/10.1103/PhysRevX.13.041045} {\bibfield  {journal} {\bibinfo  {journal} {Phys. Rev. X}\ }\textbf {\bibinfo {volume} {13}},\ \bibinfo {pages} {041045} (\bibinfo {year} {2023})}\BibitemShut {NoStop}%
\bibitem [{\citenamefont {Poboiko}\ \emph {et~al.}(2023)\citenamefont {Poboiko}, \citenamefont {P{\"o}pperl}, \citenamefont {Gornyi},\ and\ \citenamefont {Mirlin}}]{poboiko2023theory}%
  \BibitemOpen
  \bibfield  {author} {\bibinfo {author} {\bibfnamefont {I.}~\bibnamefont {Poboiko}}, \bibinfo {author} {\bibfnamefont {P.}~\bibnamefont {P{\"o}pperl}}, \bibinfo {author} {\bibfnamefont {I.~V.}\ \bibnamefont {Gornyi}},\ and\ \bibinfo {author} {\bibfnamefont {A.~D.}\ \bibnamefont {Mirlin}},\ }\bibfield  {title} {\bibinfo {title} {Theory of free fermions under random projective measurements},\ }\href {http://arxiv.org/abs/2304.03138} {\bibfield  {journal} {\bibinfo  {journal} {Phys. Rev. X}\ }\textbf {\bibinfo {volume} {13}},\ \bibinfo {pages} {041046} (\bibinfo {year} {2023})}\BibitemShut {NoStop}%
\bibitem [{\citenamefont {Fava}\ \emph {et~al.}(2024)\citenamefont {Fava}, \citenamefont {Piroli}, \citenamefont {Bernard},\ and\ \citenamefont {Nahum}}]{fava2024tractable}%
  \BibitemOpen
  \bibfield  {author} {\bibinfo {author} {\bibfnamefont {M.}~\bibnamefont {Fava}}, \bibinfo {author} {\bibfnamefont {L.}~\bibnamefont {Piroli}}, \bibinfo {author} {\bibfnamefont {D.}~\bibnamefont {Bernard}},\ and\ \bibinfo {author} {\bibfnamefont {A.}~\bibnamefont {Nahum}},\ }\bibfield  {title} {\bibinfo {title} {A tractable model of monitored fermions with conserved {$\mathrm{U}(1)$} charge},\ }\href {http://arxiv.org/abs/2407.08045} {\bibfield  {journal} {\bibinfo  {journal} {arXiv:2407.08045}\ } (\bibinfo {year} {2024})}\BibitemShut {NoStop}%
\bibitem [{\citenamefont {Majidy}\ \emph {et~al.}(2023)\citenamefont {Majidy}, \citenamefont {Agrawal}, \citenamefont {Gopalakrishnan}, \citenamefont {Potter}, \citenamefont {Vasseur},\ and\ \citenamefont {Halpern}}]{majidy2023critical}%
  \BibitemOpen
  \bibfield  {author} {\bibinfo {author} {\bibfnamefont {S.}~\bibnamefont {Majidy}}, \bibinfo {author} {\bibfnamefont {U.}~\bibnamefont {Agrawal}}, \bibinfo {author} {\bibfnamefont {S.}~\bibnamefont {Gopalakrishnan}}, \bibinfo {author} {\bibfnamefont {A.~C.}\ \bibnamefont {Potter}}, \bibinfo {author} {\bibfnamefont {R.}~\bibnamefont {Vasseur}},\ and\ \bibinfo {author} {\bibfnamefont {N.~Y.}\ \bibnamefont {Halpern}},\ }\bibfield  {title} {\bibinfo {title} {Critical phase and spin sharpening in {{SU}}(2)-symmetric monitored quantum circuits},\ }\href {https://link.aps.org/doi/10.1103/PhysRevB.108.054307} {\bibfield  {journal} {\bibinfo  {journal} {Phys. Rev. B}\ }\textbf {\bibinfo {volume} {108}},\ \bibinfo {pages} {054307} (\bibinfo {year} {2023})}\BibitemShut {NoStop}%
\bibitem [{\citenamefont {Chakraborty}\ \emph {et~al.}(2023)\citenamefont {Chakraborty}, \citenamefont {Chen}, \citenamefont {Zabalo}, \citenamefont {Wilson},\ and\ \citenamefont {Pixley}}]{chakraborty2023charge}%
  \BibitemOpen
  \bibfield  {author} {\bibinfo {author} {\bibfnamefont {A.}~\bibnamefont {Chakraborty}}, \bibinfo {author} {\bibfnamefont {K.}~\bibnamefont {Chen}}, \bibinfo {author} {\bibfnamefont {A.}~\bibnamefont {Zabalo}}, \bibinfo {author} {\bibfnamefont {J.~H.}\ \bibnamefont {Wilson}},\ and\ \bibinfo {author} {\bibfnamefont {J.~H.}\ \bibnamefont {Pixley}},\ }\bibfield  {title} {\bibinfo {title} {Charge and {{Entanglement Criticality}} in a {{U}}(1)-{{Symmetric Hybrid Circuit}} of {{Qubits}}},\ }\href {http://arxiv.org/abs/2307.13038} {\bibfield  {journal} {\bibinfo  {journal} {10.48550/arXiv.2307.13038}\ } (\bibinfo {year} {2023})}\BibitemShut {NoStop}%
\bibitem [{\citenamefont {Lavasani}\ \emph {et~al.}(2021{\natexlab{b}})\citenamefont {Lavasani}, \citenamefont {Alavirad},\ and\ \citenamefont {Barkeshli}}]{lavasani2021topological}%
  \BibitemOpen
  \bibfield  {author} {\bibinfo {author} {\bibfnamefont {A.}~\bibnamefont {Lavasani}}, \bibinfo {author} {\bibfnamefont {Y.}~\bibnamefont {Alavirad}},\ and\ \bibinfo {author} {\bibfnamefont {M.}~\bibnamefont {Barkeshli}},\ }\bibfield  {title} {\bibinfo {title} {Topological {{Order}} and {{Criticality}} in {$(2+1)\mathrm{D}$} {{Monitored Random Quantum Circuits}}},\ }\href {https://link.aps.org/doi/10.1103/PhysRevLett.127.235701} {\bibfield  {journal} {\bibinfo  {journal} {Phys. Rev. Lett.}\ }\textbf {\bibinfo {volume} {127}},\ \bibinfo {pages} {235701} (\bibinfo {year} {2021}{\natexlab{b}})}\BibitemShut {NoStop}%
\bibitem [{\citenamefont {Kells}\ \emph {et~al.}(2023)\citenamefont {Kells}, \citenamefont {Meidan},\ and\ \citenamefont {Romito}}]{kells2023topological}%
  \BibitemOpen
  \bibfield  {author} {\bibinfo {author} {\bibfnamefont {G.}~\bibnamefont {Kells}}, \bibinfo {author} {\bibfnamefont {D.}~\bibnamefont {Meidan}},\ and\ \bibinfo {author} {\bibfnamefont {A.}~\bibnamefont {Romito}},\ }\bibfield  {title} {\bibinfo {title} {Topological transitions in weakly monitored free fermions},\ }\href {https://scipost.org/10.21468/SciPostPhys.14.3.031} {\bibfield  {journal} {\bibinfo  {journal} {SciPost Physics}\ }\textbf {\bibinfo {volume} {14}},\ \bibinfo {pages} {031} (\bibinfo {year} {2023})}\BibitemShut {NoStop}%
\bibitem [{\citenamefont {Behrends}\ \emph {et~al.}(2024)\citenamefont {Behrends}, \citenamefont {Venn},\ and\ \citenamefont {B{\'e}ri}}]{behrends2024surface}%
  \BibitemOpen
  \bibfield  {author} {\bibinfo {author} {\bibfnamefont {J.}~\bibnamefont {Behrends}}, \bibinfo {author} {\bibfnamefont {F.}~\bibnamefont {Venn}},\ and\ \bibinfo {author} {\bibfnamefont {B.}~\bibnamefont {B{\'e}ri}},\ }\bibfield  {title} {\bibinfo {title} {Surface codes, quantum circuits, and entanglement phases},\ }\href {https://link.aps.org/doi/10.1103/PhysRevResearch.6.013137} {\bibfield  {journal} {\bibinfo  {journal} {Phys. Rev. Res.}\ }\textbf {\bibinfo {volume} {6}},\ \bibinfo {pages} {013137} (\bibinfo {year} {2024})}\BibitemShut {NoStop}%
\bibitem [{\citenamefont {Nahum}\ and\ \citenamefont {Skinner}(2020)}]{nahum2020entanglement}%
  \BibitemOpen
  \bibfield  {author} {\bibinfo {author} {\bibfnamefont {A.}~\bibnamefont {Nahum}}\ and\ \bibinfo {author} {\bibfnamefont {B.}~\bibnamefont {Skinner}},\ }\bibfield  {title} {\bibinfo {title} {Entanglement and dynamics of diffusion-annihilation processes with {{Majorana}} defects},\ }\href {https://link.aps.org/doi/10.1103/PhysRevResearch.2.023288} {\bibfield  {journal} {\bibinfo  {journal} {Phys. Rev. Res.}\ }\textbf {\bibinfo {volume} {2}},\ \bibinfo {pages} {023288} (\bibinfo {year} {2020})}\BibitemShut {NoStop}%
\bibitem [{\citenamefont {Cao}\ \emph {et~al.}(2019)\citenamefont {Cao}, \citenamefont {Tilloy},\ and\ \citenamefont {De~Luca}}]{cao2019entanglement}%
  \BibitemOpen
  \bibfield  {author} {\bibinfo {author} {\bibfnamefont {X.}~\bibnamefont {Cao}}, \bibinfo {author} {\bibfnamefont {A.}~\bibnamefont {Tilloy}},\ and\ \bibinfo {author} {\bibfnamefont {A.}~\bibnamefont {De~Luca}},\ }\bibfield  {title} {\bibinfo {title} {Entanglement in a fermion chain under continuous monitoring},\ }\href {https://scipost.org/SciPostPhys.7.2.024} {\bibfield  {journal} {\bibinfo  {journal} {SciPost Physics}\ }\textbf {\bibinfo {volume} {7}},\ \bibinfo {pages} {024} (\bibinfo {year} {2019})}\BibitemShut {NoStop}%
\bibitem [{\citenamefont {Alberton}\ \emph {et~al.}(2021)\citenamefont {Alberton}, \citenamefont {Buchhold},\ and\ \citenamefont {Diehl}}]{alberton2021entanglement}%
  \BibitemOpen
  \bibfield  {author} {\bibinfo {author} {\bibfnamefont {O.}~\bibnamefont {Alberton}}, \bibinfo {author} {\bibfnamefont {M.}~\bibnamefont {Buchhold}},\ and\ \bibinfo {author} {\bibfnamefont {S.}~\bibnamefont {Diehl}},\ }\bibfield  {title} {\bibinfo {title} {Entanglement {{Transition}} in a {{Monitored Free-Fermion Chain}}: {{From Extended Criticality}} to {{Area Law}}},\ }\href {https://link.aps.org/doi/10.1103/PhysRevLett.126.170602} {\bibfield  {journal} {\bibinfo  {journal} {Phys. Rev. Lett.}\ }\textbf {\bibinfo {volume} {126}},\ \bibinfo {pages} {170602} (\bibinfo {year} {2021})}\BibitemShut {NoStop}%
\bibitem [{\citenamefont {Buchhold}\ \emph {et~al.}(2021)\citenamefont {Buchhold}, \citenamefont {Minoguchi}, \citenamefont {Altland},\ and\ \citenamefont {Diehl}}]{buchhold2021effective}%
  \BibitemOpen
  \bibfield  {author} {\bibinfo {author} {\bibfnamefont {M.}~\bibnamefont {Buchhold}}, \bibinfo {author} {\bibfnamefont {Y.}~\bibnamefont {Minoguchi}}, \bibinfo {author} {\bibfnamefont {A.}~\bibnamefont {Altland}},\ and\ \bibinfo {author} {\bibfnamefont {S.}~\bibnamefont {Diehl}},\ }\bibfield  {title} {\bibinfo {title} {Effective {{Theory}} for the {{Measurement-Induced Phase Transition}} of {{Dirac Fermions}}},\ }\href {https://link.aps.org/doi/10.1103/PhysRevX.11.041004} {\bibfield  {journal} {\bibinfo  {journal} {Phys. Rev. X}\ }\textbf {\bibinfo {volume} {11}},\ \bibinfo {pages} {041004} (\bibinfo {year} {2021})}\BibitemShut {NoStop}%
\bibitem [{\citenamefont {Poboiko}\ \emph {et~al.}(2024)\citenamefont {Poboiko}, \citenamefont {Gornyi},\ and\ \citenamefont {Mirlin}}]{poboiko2024measurementinduced}%
  \BibitemOpen
  \bibfield  {author} {\bibinfo {author} {\bibfnamefont {I.}~\bibnamefont {Poboiko}}, \bibinfo {author} {\bibfnamefont {I.~V.}\ \bibnamefont {Gornyi}},\ and\ \bibinfo {author} {\bibfnamefont {A.~D.}\ \bibnamefont {Mirlin}},\ }\bibfield  {title} {\bibinfo {title} {Measurement-{{Induced Phase Transition}} for {{Free Fermions}} above {{One Dimension}}},\ }\href {https://link.aps.org/doi/10.1103/PhysRevLett.132.110403} {\bibfield  {journal} {\bibinfo  {journal} {Phys. Rev. Lett.}\ }\textbf {\bibinfo {volume} {132}},\ \bibinfo {pages} {110403} (\bibinfo {year} {2024})}\BibitemShut {NoStop}%
\bibitem [{\citenamefont {Chahine}\ and\ \citenamefont {Buchhold}(2024)}]{chahine2024entanglement}%
  \BibitemOpen
  \bibfield  {author} {\bibinfo {author} {\bibfnamefont {K.}~\bibnamefont {Chahine}}\ and\ \bibinfo {author} {\bibfnamefont {M.}~\bibnamefont {Buchhold}},\ }\bibfield  {title} {\bibinfo {title} {Entanglement phases, localization, and multifractality of monitored free fermions in two dimensions},\ }\href {https://link.aps.org/doi/10.1103/PhysRevB.110.054313} {\bibfield  {journal} {\bibinfo  {journal} {Phys. Rev. B}\ }\textbf {\bibinfo {volume} {110}},\ \bibinfo {pages} {054313} (\bibinfo {year} {2024})}\BibitemShut {NoStop}%
\bibitem [{\citenamefont {Fidkowski}\ \emph {et~al.}(2021)\citenamefont {Fidkowski}, \citenamefont {Haah},\ and\ \citenamefont {Hastings}}]{fidkowski2021howa}%
  \BibitemOpen
  \bibfield  {author} {\bibinfo {author} {\bibfnamefont {L.}~\bibnamefont {Fidkowski}}, \bibinfo {author} {\bibfnamefont {J.}~\bibnamefont {Haah}},\ and\ \bibinfo {author} {\bibfnamefont {M.~B.}\ \bibnamefont {Hastings}},\ }\bibfield  {title} {\bibinfo {title} {How {{Dynamical Quantum Memories Forget}}},\ }\href {https://quantum-journal.org/papers/q-2021-01-17-382/} {\bibfield  {journal} {\bibinfo  {journal} {Quantum}\ }\textbf {\bibinfo {volume} {5}},\ \bibinfo {pages} {382} (\bibinfo {year} {2021})}\BibitemShut {NoStop}%
\bibitem [{\citenamefont {Merritt}\ and\ \citenamefont {Fidkowski}(2023)}]{merritt2023entanglement}%
  \BibitemOpen
  \bibfield  {author} {\bibinfo {author} {\bibfnamefont {J.}~\bibnamefont {Merritt}}\ and\ \bibinfo {author} {\bibfnamefont {L.}~\bibnamefont {Fidkowski}},\ }\bibfield  {title} {\bibinfo {title} {Entanglement transitions with free fermions},\ }\href {https://link.aps.org/doi/10.1103/PhysRevB.107.064303} {\bibfield  {journal} {\bibinfo  {journal} {Phys. Rev. B}\ }\textbf {\bibinfo {volume} {107}},\ \bibinfo {pages} {064303} (\bibinfo {year} {2023})}\BibitemShut {NoStop}%
\bibitem [{\citenamefont {Chen}\ \emph {et~al.}(2020)\citenamefont {Chen}, \citenamefont {Li}, \citenamefont {Fisher},\ and\ \citenamefont {Lucas}}]{chen2020emergent}%
  \BibitemOpen
  \bibfield  {author} {\bibinfo {author} {\bibfnamefont {X.}~\bibnamefont {Chen}}, \bibinfo {author} {\bibfnamefont {Y.}~\bibnamefont {Li}}, \bibinfo {author} {\bibfnamefont {M.~P.~A.}\ \bibnamefont {Fisher}},\ and\ \bibinfo {author} {\bibfnamefont {A.}~\bibnamefont {Lucas}},\ }\bibfield  {title} {\bibinfo {title} {Emergent conformal symmetry in nonunitary random dynamics of free fermions},\ }\href {https://link.aps.org/doi/10.1103/PhysRevResearch.2.033017} {\bibfield  {journal} {\bibinfo  {journal} {Phys. Rev. Res.}\ }\textbf {\bibinfo {volume} {2}},\ \bibinfo {pages} {033017} (\bibinfo {year} {2020})}\BibitemShut {NoStop}%
\bibitem [{\citenamefont {Tang}\ \emph {et~al.}(2021)\citenamefont {Tang}, \citenamefont {Chen},\ and\ \citenamefont {Zhu}}]{tang2021quantum}%
  \BibitemOpen
  \bibfield  {author} {\bibinfo {author} {\bibfnamefont {Q.}~\bibnamefont {Tang}}, \bibinfo {author} {\bibfnamefont {X.}~\bibnamefont {Chen}},\ and\ \bibinfo {author} {\bibfnamefont {W.}~\bibnamefont {Zhu}},\ }\bibfield  {title} {\bibinfo {title} {Quantum criticality in the nonunitary dynamics of {$(2+1)$}-dimensional free fermions},\ }\href {https://link.aps.org/doi/10.1103/PhysRevB.103.174303} {\bibfield  {journal} {\bibinfo  {journal} {Phys. Rev. B}\ }\textbf {\bibinfo {volume} {103}},\ \bibinfo {pages} {174303} (\bibinfo {year} {2021})}\BibitemShut {NoStop}%
\bibitem [{\citenamefont {Guo}\ \emph {et~al.}(2024)\citenamefont {Guo}, \citenamefont {Foster}, \citenamefont {Jian},\ and\ \citenamefont {Ludwig}}]{guo2024field}%
  \BibitemOpen
  \bibfield  {author} {\bibinfo {author} {\bibfnamefont {H.}~\bibnamefont {Guo}}, \bibinfo {author} {\bibfnamefont {M.~S.}\ \bibnamefont {Foster}}, \bibinfo {author} {\bibfnamefont {C.-M.}\ \bibnamefont {Jian}},\ and\ \bibinfo {author} {\bibfnamefont {A.~W.~W.}\ \bibnamefont {Ludwig}},\ }\bibfield  {title} {\bibinfo {title} {Field theory of monitored, interacting fermion dynamics with charge conservation},\ }\href {https://arxiv.org/abs/v1} {\bibfield  {journal} {\bibinfo  {journal} {arXiv:2410.07317}\ } (\bibinfo {year} {2024})}\BibitemShut {NoStop}%
\bibitem [{\citenamefont {{Poboiko}}\ \emph {et~al.}(2025)\citenamefont {{Poboiko}}, \citenamefont {{P{\"o}pperl}}, \citenamefont {{Gornyi}},\ and\ \citenamefont {{Mirlin}}}]{MirlinInteracting2024}%
  \BibitemOpen
  \bibfield  {author} {\bibinfo {author} {\bibfnamefont {I.}~\bibnamefont {{Poboiko}}}, \bibinfo {author} {\bibfnamefont {P.}~\bibnamefont {{P{\"o}pperl}}}, \bibinfo {author} {\bibfnamefont {I.~V.}\ \bibnamefont {{Gornyi}}},\ and\ \bibinfo {author} {\bibfnamefont {A.~D.}\ \bibnamefont {{Mirlin}}},\ }\bibfield  {title} {\bibinfo {title} {{Measurement-induced transitions for interacting fermions}},\ }\href {https://doi.org/10.1103/PhysRevB.111.024204} {\bibfield  {journal} {\bibinfo  {journal} {\prb}\ }\textbf {\bibinfo {volume} {111}},\ \bibinfo {eid} {024204} (\bibinfo {year} {2025})},\ \Eprint {https://arxiv.org/abs/2410.07334} {arXiv:2410.07334 [quant-ph]} \BibitemShut {NoStop}%
\bibitem [{\citenamefont {Altland}\ and\ \citenamefont {Zirnbauer}(1997)}]{altland1997nonstandard}%
  \BibitemOpen
  \bibfield  {author} {\bibinfo {author} {\bibfnamefont {A.}~\bibnamefont {Altland}}\ and\ \bibinfo {author} {\bibfnamefont {M.~R.}\ \bibnamefont {Zirnbauer}},\ }\bibfield  {title} {\bibinfo {title} {Nonstandard symmetry classes in mesoscopic normal-superconducting hybrid structures},\ }\href {https://link.aps.org/doi/10.1103/PhysRevB.55.1142} {\bibfield  {journal} {\bibinfo  {journal} {Phys. Rev. B}\ }\textbf {\bibinfo {volume} {55}},\ \bibinfo {pages} {1142} (\bibinfo {year} {1997})}\BibitemShut {NoStop}%
\bibitem [{\citenamefont {Ludwig}(2016)}]{ludwig2016topological}%
  \BibitemOpen
  \bibfield  {author} {\bibinfo {author} {\bibfnamefont {A.~W.~W.}\ \bibnamefont {Ludwig}},\ }\bibfield  {title} {\bibinfo {title} {Topological phases: {{Classification}} of topological insulators and superconductors of non-interacting {{Fermions}}, and beyond},\ }\href {http://arxiv.org/abs/1512.08882} {\bibfield  {journal} {\bibinfo  {journal} {Phys. Scr.}\ }\textbf {\bibinfo {volume} {T168}},\ \bibinfo {pages} {014001} (\bibinfo {year} {2016})}\BibitemShut {NoStop}%
\bibitem [{\citenamefont {Qi}\ and\ \citenamefont {Zhang}(2011)}]{qi2011topological}%
  \BibitemOpen
  \bibfield  {author} {\bibinfo {author} {\bibfnamefont {X.-L.}\ \bibnamefont {Qi}}\ and\ \bibinfo {author} {\bibfnamefont {S.-C.}\ \bibnamefont {Zhang}},\ }\bibfield  {title} {\bibinfo {title} {Topological insulators and superconductors},\ }\href {https://link.aps.org/doi/10.1103/RevModPhys.83.1057} {\bibfield  {journal} {\bibinfo  {journal} {Rev. Mod. Phys.}\ }\textbf {\bibinfo {volume} {83}},\ \bibinfo {pages} {1057} (\bibinfo {year} {2011})}\BibitemShut {NoStop}%
\bibitem [{\citenamefont {Chen}\ and\ \citenamefont {Vidal}(2014)}]{chen2014entanglement}%
  \BibitemOpen
  \bibfield  {author} {\bibinfo {author} {\bibfnamefont {Y.}~\bibnamefont {Chen}}\ and\ \bibinfo {author} {\bibfnamefont {G.}~\bibnamefont {Vidal}},\ }\bibfield  {title} {\bibinfo {title} {Entanglement contour},\ }\href {https://dx.doi.org/10.1088/1742-5468/2014/10/P10011} {\bibfield  {journal} {\bibinfo  {journal} {J. Stat. Mech.}\ }\textbf {\bibinfo {volume} {2014}},\ \bibinfo {pages} {P10011} (\bibinfo {year} {2014})}\BibitemShut {NoStop}%
\bibitem [{\citenamefont {Alicea}\ \emph {et~al.}(2011)\citenamefont {Alicea}, \citenamefont {Oreg}, \citenamefont {Refael}, \citenamefont {{von Oppen}},\ and\ \citenamefont {Fisher}}]{alicea2011nonabelian}%
  \BibitemOpen
  \bibfield  {author} {\bibinfo {author} {\bibfnamefont {J.}~\bibnamefont {Alicea}}, \bibinfo {author} {\bibfnamefont {Y.}~\bibnamefont {Oreg}}, \bibinfo {author} {\bibfnamefont {G.}~\bibnamefont {Refael}}, \bibinfo {author} {\bibfnamefont {F.}~\bibnamefont {{von Oppen}}},\ and\ \bibinfo {author} {\bibfnamefont {M.~P.~A.}\ \bibnamefont {Fisher}},\ }\bibfield  {title} {\bibinfo {title} {Non-{{Abelian}} statistics and topological quantum information processing in {{1D}} wire networks},\ }\href {https://www.nature.com/articles/nphys1915} {\bibfield  {journal} {\bibinfo  {journal} {Nature Physics}\ }\textbf {\bibinfo {volume} {7}},\ \bibinfo {pages} {412} (\bibinfo {year} {2011})}\BibitemShut {NoStop}%
\bibitem [{\citenamefont {Karzig}\ \emph {et~al.}(2016)\citenamefont {Karzig}, \citenamefont {Oreg}, \citenamefont {Refael},\ and\ \citenamefont {Freedman}}]{karzig2016universal}%
  \BibitemOpen
  \bibfield  {author} {\bibinfo {author} {\bibfnamefont {T.}~\bibnamefont {Karzig}}, \bibinfo {author} {\bibfnamefont {Y.}~\bibnamefont {Oreg}}, \bibinfo {author} {\bibfnamefont {G.}~\bibnamefont {Refael}},\ and\ \bibinfo {author} {\bibfnamefont {M.~H.}\ \bibnamefont {Freedman}},\ }\bibfield  {title} {\bibinfo {title} {Universal {{Geometric Path}} to a {{Robust Majorana Magic Gate}}},\ }\href {https://link.aps.org/doi/10.1103/PhysRevX.6.031019} {\bibfield  {journal} {\bibinfo  {journal} {Phys. Rev. X}\ }\textbf {\bibinfo {volume} {6}},\ \bibinfo {pages} {031019} (\bibinfo {year} {2016})}\BibitemShut {NoStop}%
\bibitem [{\citenamefont {Nielsen}\ and\ \citenamefont {Chuang}(2012)}]{nielsen2012quantum}%
  \BibitemOpen
  \bibfield  {author} {\bibinfo {author} {\bibfnamefont {M.~A.}\ \bibnamefont {Nielsen}}\ and\ \bibinfo {author} {\bibfnamefont {I.~L.}\ \bibnamefont {Chuang}},\ }\href {https://www.cambridge.org/core/product/identifier/9780511976667/type/book} {\emph {\bibinfo {title} {Quantum {{Computation}} and {{Quantum Information}}: 10th {{Anniversary Edition}}}}},\ \bibinfo {edition} {1st}\ ed.\ (\bibinfo  {publisher} {Cambridge University Press},\ \bibinfo {year} {2012})\BibitemShut {NoStop}%
\bibitem [{\citenamefont {Zatelli}\ \emph {et~al.}(2024)\citenamefont {Zatelli}, \citenamefont {{van Driel}}, \citenamefont {Xu}, \citenamefont {Wang}, \citenamefont {Liu}, \citenamefont {Bordin}, \citenamefont {Roovers}, \citenamefont {Mazur}, \citenamefont {{van Loo}}, \citenamefont {Wolff}, \citenamefont {Bozkurt}, \citenamefont {Badawy}, \citenamefont {Gazibegovic}, \citenamefont {Bakkers}, \citenamefont {Wimmer}, \citenamefont {Kouwenhoven},\ and\ \citenamefont {Dvir}}]{zatelli2024robust}%
  \BibitemOpen
  \bibfield  {author} {\bibinfo {author} {\bibfnamefont {F.}~\bibnamefont {Zatelli}}, \bibinfo {author} {\bibfnamefont {D.}~\bibnamefont {{van Driel}}}, \bibinfo {author} {\bibfnamefont {D.}~\bibnamefont {Xu}}, \bibinfo {author} {\bibfnamefont {G.}~\bibnamefont {Wang}}, \bibinfo {author} {\bibfnamefont {C.-X.}\ \bibnamefont {Liu}}, \bibinfo {author} {\bibfnamefont {A.}~\bibnamefont {Bordin}}, \bibinfo {author} {\bibfnamefont {B.}~\bibnamefont {Roovers}}, \bibinfo {author} {\bibfnamefont {G.~P.}\ \bibnamefont {Mazur}}, \bibinfo {author} {\bibfnamefont {N.}~\bibnamefont {{van Loo}}}, \bibinfo {author} {\bibfnamefont {J.~C.}\ \bibnamefont {Wolff}}, \bibinfo {author} {\bibfnamefont {A.~M.}\ \bibnamefont {Bozkurt}}, \bibinfo {author} {\bibfnamefont {G.}~\bibnamefont {Badawy}}, \bibinfo {author} {\bibfnamefont {S.}~\bibnamefont {Gazibegovic}}, \bibinfo {author} {\bibfnamefont {E.~P. A.~M.}\ \bibnamefont {Bakkers}}, \bibinfo {author} {\bibfnamefont {M.}~\bibnamefont {Wimmer}}, \bibinfo {author} {\bibfnamefont {L.~P.}\ \bibnamefont {Kouwenhoven}},\ and\ \bibinfo {author} {\bibfnamefont {T.}~\bibnamefont {Dvir}},\ }\bibfield  {title} {\bibinfo {title} {Robust poor man's {{Majorana}} zero modes using {{Yu-Shiba-Rusinov}} states},\ }\href {https://www.nature.com/articles/s41467-024-52066-2} {\bibfield  {journal} {\bibinfo  {journal} {Nat Commun}\ }\textbf {\bibinfo {volume} {15}},\ \bibinfo {pages} {7933} (\bibinfo {year} {2024})}\BibitemShut {NoStop}%
\bibitem [{\citenamefont {Bravyi}(2005)}]{bravyi2005lagrangian}%
  \BibitemOpen
  \bibfield  {author} {\bibinfo {author} {\bibfnamefont {S.}~\bibnamefont {Bravyi}},\ }\bibfield  {title} {\bibinfo {title} {Lagrangian representation for fermionic linear optics},\ }\href {https://dl.acm.org/doi/abs/10.5555/2011637.2011640} {\bibfield  {journal} {\bibinfo  {journal} {Quantum Info. Comput.}\ }\textbf {\bibinfo {volume} {5}},\ \bibinfo {pages} {216} (\bibinfo {year} {2005})}\BibitemShut {NoStop}%
\bibitem [{\citenamefont {Bravyi}\ and\ \citenamefont {Gosset}(2017)}]{bravyi2017complexity}%
  \BibitemOpen
  \bibfield  {author} {\bibinfo {author} {\bibfnamefont {S.}~\bibnamefont {Bravyi}}\ and\ \bibinfo {author} {\bibfnamefont {D.}~\bibnamefont {Gosset}},\ }\bibfield  {title} {\bibinfo {title} {Complexity of {{Quantum Impurity Problems}}},\ }\href {https://doi.org/10.1007/s00220-017-2976-9} {\bibfield  {journal} {\bibinfo  {journal} {Commun. Math. Phys.}\ }\textbf {\bibinfo {volume} {356}},\ \bibinfo {pages} {451} (\bibinfo {year} {2017})}\BibitemShut {NoStop}%
\bibitem [{\citenamefont {Choi}(1975)}]{choi1975completely}%
  \BibitemOpen
  \bibfield  {author} {\bibinfo {author} {\bibfnamefont {M.-D.}\ \bibnamefont {Choi}},\ }\bibfield  {title} {\bibinfo {title} {Completely positive linear maps on complex matrices},\ }\href {https://www.sciencedirect.com/science/article/pii/0024379575900750} {\bibfield  {journal} {\bibinfo  {journal} {Linear Algebra and its Applications}\ }\textbf {\bibinfo {volume} {10}},\ \bibinfo {pages} {285} (\bibinfo {year} {1975})}\BibitemShut {NoStop}%
\bibitem [{\citenamefont {Jamio{\l}kowski}(1972)}]{jamiolkowski1972linear}%
  \BibitemOpen
  \bibfield  {author} {\bibinfo {author} {\bibfnamefont {A.}~\bibnamefont {Jamio{\l}kowski}},\ }\bibfield  {title} {\bibinfo {title} {Linear transformations which preserve trace and positive semidefiniteness of operators},\ }\href {https://www.sciencedirect.com/science/article/pii/0034487772900110} {\bibfield  {journal} {\bibinfo  {journal} {Reports on Mathematical Physics}\ }\textbf {\bibinfo {volume} {3}},\ \bibinfo {pages} {275} (\bibinfo {year} {1972})}\BibitemShut {NoStop}%
\bibitem [{\citenamefont {Ericsson}\ and\ \citenamefont {{Centro di Ricerca Matematica Ennio De Giorgi}}(2008)}]{ericsson2008quantum}%
  \BibitemOpen
  \bibinfo {editor} {\bibfnamefont {M.}~\bibnamefont {Ericsson}}\ and\ \bibinfo {editor} {\bibnamefont {{Centro di Ricerca Matematica Ennio De Giorgi}}},\ eds.,\ \href@noop {} {\emph {\bibinfo {title} {Quantum Information and Many Body Quantum Systems: Proceedings}}},\ \bibinfo {series} {Centro Di {{Ricerca Matematica Ennio De Giorgi}} / {{CRM}} Series}\ No.~\bibinfo {number} {8}\ (\bibinfo  {publisher} {Ed. della Normale},\ \bibinfo {address} {Pisa},\ \bibinfo {year} {2008})\BibitemShut {NoStop}%
\bibitem [{\citenamefont {Kraus}\ \emph {et~al.}(2010)\citenamefont {Kraus}, \citenamefont {Schuch}, \citenamefont {Verstraete},\ and\ \citenamefont {Cirac}}]{kraus2010fermionic}%
  \BibitemOpen
  \bibfield  {author} {\bibinfo {author} {\bibfnamefont {C.~V.}\ \bibnamefont {Kraus}}, \bibinfo {author} {\bibfnamefont {N.}~\bibnamefont {Schuch}}, \bibinfo {author} {\bibfnamefont {F.}~\bibnamefont {Verstraete}},\ and\ \bibinfo {author} {\bibfnamefont {J.~I.}\ \bibnamefont {Cirac}},\ }\bibfield  {title} {\bibinfo {title} {Fermionic projected entangled pair states},\ }\href {https://link.aps.org/doi/10.1103/PhysRevA.81.052338} {\bibfield  {journal} {\bibinfo  {journal} {Phys. Rev. A}\ }\textbf {\bibinfo {volume} {81}},\ \bibinfo {pages} {052338} (\bibinfo {year} {2010})}\BibitemShut {NoStop}%
\bibitem [{\citenamefont {Chalker}\ and\ \citenamefont {Coddington}(1988)}]{chalker1988percolation}%
  \BibitemOpen
  \bibfield  {author} {\bibinfo {author} {\bibfnamefont {J.~T.}\ \bibnamefont {Chalker}}\ and\ \bibinfo {author} {\bibfnamefont {P.~D.}\ \bibnamefont {Coddington}},\ }\bibfield  {title} {\bibinfo {title} {Percolation, quantum tunnelling and the integer {{Hall}} effect},\ }\href {https://dx.doi.org/10.1088/0022-3719/21/14/008} {\bibfield  {journal} {\bibinfo  {journal} {J. Phys. C: Solid State Phys.}\ }\textbf {\bibinfo {volume} {21}},\ \bibinfo {pages} {2665} (\bibinfo {year} {1988})}\BibitemShut {NoStop}%
\bibitem [{\citenamefont {Brouwer}\ \emph {et~al.}(2000)\citenamefont {Brouwer}, \citenamefont {Furusaki}, \citenamefont {Gruzberg},\ and\ \citenamefont {Mudry}}]{brouwer2000localization}%
  \BibitemOpen
  \bibfield  {author} {\bibinfo {author} {\bibfnamefont {P.~W.}\ \bibnamefont {Brouwer}}, \bibinfo {author} {\bibfnamefont {A.}~\bibnamefont {Furusaki}}, \bibinfo {author} {\bibfnamefont {I.~A.}\ \bibnamefont {Gruzberg}},\ and\ \bibinfo {author} {\bibfnamefont {C.}~\bibnamefont {Mudry}},\ }\bibfield  {title} {\bibinfo {title} {Localization and {{Delocalization}} in {{Dirty Superconducting Wires}}},\ }\href {https://link.aps.org/doi/10.1103/PhysRevLett.85.1064} {\bibfield  {journal} {\bibinfo  {journal} {Phys. Rev. Lett.}\ }\textbf {\bibinfo {volume} {85}},\ \bibinfo {pages} {1064} (\bibinfo {year} {2000})}\BibitemShut {NoStop}%
\bibitem [{\citenamefont {Brouwer}\ \emph {et~al.}(2006)\citenamefont {Brouwer}, \citenamefont {Furusaki}, \citenamefont {Mudry},\ and\ \citenamefont {Ryu}}]{brouwer2006disorderinduced}%
  \BibitemOpen
  \bibfield  {author} {\bibinfo {author} {\bibfnamefont {P.~W.}\ \bibnamefont {Brouwer}}, \bibinfo {author} {\bibfnamefont {A.}~\bibnamefont {Furusaki}}, \bibinfo {author} {\bibfnamefont {C.}~\bibnamefont {Mudry}},\ and\ \bibinfo {author} {\bibfnamefont {S.}~\bibnamefont {Ryu}},\ }\bibfield  {title} {\bibinfo {title} {Disorder-induced critical phenomena--new universality classes in {{Anderson}} localization},\ }\href {http://arxiv.org/abs/cond-mat/0511622} {\bibfield  {journal} {\bibinfo  {journal} {arXiv:cond-mat/0511622}\ } (\bibinfo {year} {2006})}\BibitemShut {NoStop}%
\bibitem [{\citenamefont {Ludwig}\ \emph {et~al.}(2013)\citenamefont {Ludwig}, \citenamefont {{Schulz-Baldes}},\ and\ \citenamefont {Stolz}}]{ludwig2013lyapunov}%
  \BibitemOpen
  \bibfield  {author} {\bibinfo {author} {\bibfnamefont {A.~W.~W.}\ \bibnamefont {Ludwig}}, \bibinfo {author} {\bibfnamefont {H.}~\bibnamefont {{Schulz-Baldes}}},\ and\ \bibinfo {author} {\bibfnamefont {M.}~\bibnamefont {Stolz}},\ }\bibfield  {title} {\bibinfo {title} {Lyapunov {{Spectra}} for {{All Ten Symmetry Classes}} of {{Quasi-one-dimensional Disordered Systems}} of {{Non-interacting Fermions}}},\ }\href {https://doi.org/10.1007/s10955-013-0764-2} {\bibfield  {journal} {\bibinfo  {journal} {J Stat Phys}\ }\textbf {\bibinfo {volume} {152}},\ \bibinfo {pages} {275} (\bibinfo {year} {2013})}\BibitemShut {NoStop}%
\end{thebibliography}%
\renewcommand{\addcontentsline}[3]{\oldaddcontentsline{#1}{#2}{#3}}

\clearpage
\appendix
\renewcommand{\thesection}{\Roman{section}}
\vspace{3cm}
\onecolumngrid
\twocolumngrid
\setcounter{page}{1}
\setcounter{secnumdepth}{3}
\setcounter{equation}{0}
\setcounter{figure}{0}
\renewcommand{\theequation}{S-\thesection.\arabic{equation}}

\renewcommand{\thefigure}{S\arabic{figure}}
\renewcommand\figurename{Supplementary Figure}
\renewcommand\tablename{Supplementary Table}
\tableofcontents

\section{Free-fermion monitored circuit}\label{sec:Gaussian}
In this section, we provide a brief review of the free-fermion monitored circuit. In the following, we will refer to free-fermion states as Gaussian states and the operators that keep the many-body fermionic states ``free" as Gaussian operators. Here, the Gaussianity refers to the fact that, in the free-fermion setting, the system's density operator and the operations on them all take a Gaussian form, i.e., exponentiated fermion bilinears.
We will demonstrate, through a concrete example, that a free-fermion circuit consisting of Gaussian operators can be efficiently simulated in a covariance matrix formalism.
For a more rigorous treatment, we refer the interested readers to Refs.~\cite{bravyi2005lagrangian,bravyi2017complexity}. 

\subsection{Efficient simulation of Gaussian circuits}
We start with a generic Gaussian state in its density matrix form $\rho$ and Gaussian operators $\mathcal{O}$, both constructed in the Majorana basis $\hat{\gamma}_i$,  in the following form
\begin{equation}\label{eq:gaussian}
    \rho\propto e^{\sum_{i,j} \ii\hat{\gamma}_i H_{ij} \hat{\gamma}_j },
    \mathcal{O}\propto e^{\sum_{i,j} \hat{\gamma}_i O_{ij} \hat{\gamma}_j},
\end{equation}
where $H$ and $O$ are real and complex antisymmetric matrix (with $H_{ij},\Im(O_{ij})\rightarrow\pm\infty$ allowed), respectively. Here, the Gaussian operator $\mathcal{O}$ includes the unitary gates and the Kraus operators associated with measurements.

Our goal is to show how the state evolution of $\rho$ under $\mathcal{O}$, i.e., $\mathcal{O}\rho\mathcal{O}^\dagger$, can be efficiently simulated using covariance matrix formalism, which is also the numerical method in the main text.

To this end, we first define the covariance matrix $\Gamma_\rho$ for the Gaussian state $\rho$ as
\begin{equation}\label{eq:Gamma} 
    \left[ \Gamma_{\rho} \right]_{ij}=\frac{\ii}{2}\text{tr}( [\hat{\gamma}_i,\hat{\gamma}_j]\rho).
\end{equation}
The covariance matrix for the Gaussian operator $\mathcal{O}$ needs a special treatment following the Choi-Jamio\l kowski isomorphism~\cite{choi1975completely,jamiolkowski1972linear,guo2024field} which generalizes the original Hilbert space spanned by $\hat{\gamma}_i$ to a doubled Hilbert space spanned by  $\hat{\eta}_i$ and $\hat{\xi}_i$ (two sets of Majorana fermions in a different space from $\hat{\gamma}_i$) in the following form
\begin{equation}\label{eq:blocks}
    \Gamma_{\mathcal{O}}=\begin{pmatrix}
        \Gamma_{\mathcal{O},\eta\eta} & \Gamma_{\mathcal{O},\eta\xi}\\
        \Gamma_{\mathcal{O},\xi\eta} & \Gamma_{\mathcal{O},\xi\xi}
    \end{pmatrix},
\end{equation}
where each block is defined as
\begin{equation}
    \begin{split}
        &\left[ \Gamma_{\mathcal{O},\alpha\beta} \right]_{ij}= \\
        & \frac{1}{2}\text{tr}\left( e^{\sum_{i,j} \hat{\eta}_i O_{ij} \hat{\eta}_j} \prod_{i}\frac{1+\ii\hat{\eta}_i\hat{\xi}_i}{2} e^{\sum_{i,j} \hat{\eta}_i O_{ji}^* \hat{\eta}_j} ~\ii[\hat{\alpha}_i,\hat{\beta}_j] \right),
    \end{split}
\end{equation}
with $\alpha,\beta$ taking $\eta$ or $\xi$ for the corresponding blocks in Eq.~\eqref{eq:blocks}. 

With these two covariance matrices representation $\Gamma_\rho$ and $\Gamma_{\mathcal{O}}$ for the state $\rho$ and operator $\mathcal{O}$, we can readily obtain the resulting state $\rho^\prime = \mathcal{O}\rho\mathcal{O}^\dagger$ in the covariance matrices representation $\Gamma_{\rho^\prime}$ using the following contraction formula~\cite{bravyi2005lagrangian}
\begin{equation}\label{eq:contraction}
    \Gamma_{\rho^\prime}=\Gamma_{\mathcal{O},\xi\xi}+\Gamma_{\mathcal{O},\xi\eta}\left(\Gamma_{\mathcal{O},\eta\eta}+\Gamma_{\rho}^{-1}\right)^{-1}\Gamma_{\mathcal{O},\xi\eta}^\intercal.
\end{equation}

In numerical simulation, we initialize a random pure Gaussian state $\Gamma_{\rho_0}$ in a $L$-site Majorana chain as
\begin{equation}\label{eq:random_initial}
    \Gamma_{\rho_0}=V \bigoplus_{i=1}^{L/2} \begin{pmatrix}
        0   & \Pi_i\\
        -\Pi_i & 0
    \end{pmatrix} V^\intercal,
\end{equation}
with a random matrix $V\in\text{SO}(2L)$ and $\Pi_i=\pm1$, and repeat the contraction formula Eq.~\eqref{eq:contraction} for various $\mathcal{O}$ to simulate the circuit in $O(L^3)$ complexity.

For the class-DIII monitored circuit, we choose all $\Pi_i=1$.
For the class-A monitored circuit, due to the U(1) charge conservation, we park ourselves at the half-filling sector by choosing a random profile of $\Pi_i=\pm1$ subject to $\sum_i\Pi_i=0$.

After the contraction in Eq.~\eqref{eq:contraction}, the resulting state could be slightly deviated away from a pure state as $\abs{\Pi_i}<1$ due to the numerical error.
Therefore, we purify the state once in a while by block-diagonalizing the covariance matrix into Eq.~\eqref{eq:random_initial}, and resetting all $\Pi_i$ to its nearest $\pm1$.

\subsection{Entanglement measures in the covariance matrix formalism}\label{sec:EE}
Besides the efficient simulation of the circuit in the covariance matrix formalism, all entanglement measures for a pure Gaussian state $\rho$ can be also efficiently computed in this formalism. Below, we will demonstrate how to efficiently compute the entanglement entropy (EE), entanglement contour (EC), and the mutual information (MI) for a Gaussian state encoded in the covariance matrix $\Gamma_\rho$.
We refer the interested readers to Ref.~\cite{chen2014entanglement} for a more rigorous treatment.

The von Neumann EE between a region $A$ and its complement $\bar{A}$ of state $\rho$ is obtained from its reduced density matrix $\rho_A = \tr_{\bar{A}}\rho$, whose covariance matrix representation can be obtained as  $\Gamma_A= \sum_{i,j\in A} \left[ \Gamma_\rho \right]_{i,j} \ketbra{i}{j}$.

In the covariance matrix formalism, the von Neumann EE of $\Gamma_A$ is computed as $S_A=\tr f(\Gamma_A)$ with
\begin{equation}\label{eq:f}
    f(\Gamma_A)=-\frac{\mathds{1}+\ii\Gamma_A}{2}\log\left(\frac{\mathds{1}+\ii\Gamma_A}{2}\right).
\end{equation}

The EC is a spatially-resolved EE~\cite{chen2014entanglement}, indicating the contribution of each site to the total EE, which can be obtained by taking the $i$th diagonal element of the matrix of $f(\Gamma_A)$, namely
\begin{equation}\label{eq:EC}
    s_A(i)=\left[ f(\Gamma_A) \right]_{ii},
\end{equation}
with $f(\Gamma_A)$ defined in Eq.~\eqref{eq:f}.
This definition ensures that the entanglement contour on each site $i$ sums up to the total EE of $A$, i.e., $\sum_{i\in A}s_A(i)=S_A$.
Here, we generalize the original version of the EC in Ref.~\cite{chen2014entanglement} from complex fermions to Majorana fermions.

Finally, to identify the volume-law and area-law phases, we use MI $I_{A,B}$ between two antipodal regions $A$ and $B$, each of one-quarter of the system, defined as 
\begin{equation}
    I_{A,B}=S_A+S_B-S_{A\cup B}.
\end{equation}

\section{Correspondence between a free-fermion monitored circuit and Anderson localization problem}\label{sec:correspondence}

In this section, we briefly review the correspondence between a free-fermion monitored circuit in 1+1D and the Anderson localization problem in 2 spatial dimensions. We will focus on the version of this correspondence between the 1+1D monitored circuit of free Majorana fermions without any constraints and the 2D Anderson localization problem of AZ symmetry class DIII. We will extend the discussion to class AIII and class A. For more details, we refer the interested readers to Ref.~\cite{jian2020measurementinduced}. 

We start with a generic free-fermion monitored circuit on a Majorana fermion chain without any additional symmetry, as shown in Fig.~\ref{fig:DIII}(a).
We can treat the time and spatial axis on an equal footing and view the circuit as a Gaussian tensor network~\cite{ericsson2008quantum,kraus2010fermionic} on a 2D square lattice.
Therefore, the evolution of the Gaussian state under the Gaussian operator $\mathcal{O}$ can be interpreted as the contraction of Gaussian tensors. For example, in the covariance matrix formalism, this evolution performs the contraction between a Gaussian state $\rho$ and a Gaussian operator $\mathcal{O}$ following Eq.~\eqref{eq:contraction}. In a given quantum trajectory, the quantum gates in the circuit can be viewed as the many-body transfer operator of the corresponding Gaussian tensor network. These quantum gates (or transfer operators), due to their Gaussianity, admit a ``first-quantized'' transfer matrix representation describing the transformation of Majorana fermion operators $\hat{\gamma}_i$ under $\mathcal{O}$, i.e.,
\begin{equation}
    \mathcal{O} \hat{\gamma}_i \mathcal{O}^{-1} = \sum_j \mathfrak{t}_{ij} \hat{\gamma}_j.
\end{equation} 
If $\mathcal{O}$ is a unitary gate, we know that $\mathfrak{t}$ is a real orthogonal matrix in SO$(N)$ where $N$ is the number of the Majorana modes.
However, when we include measurements, we can show that the transfer matrix $\mathfrak{t}$ become a complexified orthogonal matrix in SO$(N,\mathbb{C})$, i.e. $\mathfrak{t}$ is a complex $N\times N$ matrix satisfying $\mathfrak{t}^\mathsf{T} \mathfrak{t} = \mathds{1}$ \cite{jian2022criticality}. We caution that, when $\mathcal{O}$ is associated with a projective measurement, $\mathcal{O}^{-1}$ and, consequently, $\mathfrak{t}$ becomes singular. In this case, we should view the projective measurements as a limit of weak measurements with infinite measurement strength. For weak measurements of finite strength, the transfer matrix $\mathfrak{t}$ is non-singular. In a monitored circuit, different measurement outcomes (and different random realizations of the unitary gates) correspond to different realizations of the gates ${\cal O}$ in the circuit. Therefore, the 1+1D monitored circuit of free Majorana fermions shown in Fig. \ref{fig:DIII}(a) can be interpreted as a 2D random Gaussian tensor network whose single-particle transfer matrices are sampled from the complexified orthogonal group SO$(N,\mathbb{C})$.

A 2D Gaussian tensor network with SO$(N,\mathbb{C})$ single-particle transfer can also be interpreted as a Chalker-Coddington network model~\cite{chalker1988percolation} that describes a network of disordered elastic scatters in the AZ symmetry class DIII~\cite{jian2022criticality}. This Chalker-Coddington network model captures the Anderson localization problem in static disordered Hamiltonian of non-interacting fermions in 2 spatial dimensions in symmetry class DIII. It was shown that, in a non-interacting-fermion Hamiltonian system in symmetry class DIII, namely a system equipped with a time-reversal symmetry (that squares to $-\mathds{1}$) and a particle-hole (that squares to $\mathds{1}$), the single-particle impurity scattering can be captured by SO$(N,\mathbb{C})$ transfer matrices \cite{brouwer2000localization,brouwer2006disorderinduced,ludwig2013lyapunov}. This fact enables the aforementioned identification between the Gaussian tensor network and the Anderson localization problem.

From the complementary perspective of the monitored circuit, even if we do not impose any symmetry requirement on the Majorana chain (other than fermion parity), when we consider the evolution of the system's density matrix in a doubled Hilbert space, extra symmetries that relate the ``bra" and ``ket" copy of the Hilbert space can emerge. In the case of the monitored Majorana chain, the total emergent symmetry belongs to AZ symmetry class DIII~\cite{jian2022criticality}. The correspondence between the monitored circuit of free Majorana fermions in 1+1D and the 2D disordered Anderson localization problem in symmetry class DIII implies that the area-law entangled phases in the former setting and the disordered topological superconductors in the latter problem should share the same topological classification, which is $\mathbb{Z}_2$\cite{qi2011topological,ludwig2016topological}.

When we add an extra symmetry constraint to the monitored circuit, the group of transfer matrices changes, and so does the symmetry class. The identification with the Anderson localization problem remains correct in the respective symmetry class. When we consider a monitored circuit with U(1) charge conservation, the transfer matrix belongs to GL$(N,\mathbb{C})$, and the symmetry class becomes class AIII. Furthermore, adding an anti-unitary chiral symmetry, as we do in the main text, limits the transfer matrix to the group U$(N,N)$, resulting in symmetry class A. Concrete examples of the quantum gates and their transfer matrices in class AIII and class A are provided in Sec.~\ref{sec:AIII} and Sec.~\ref{sec:A}.

\section{Class-DIII monitored circuit}\label{sec:DIII}
In this section, we first provide a more formal treatment of the two types of operators in the class-DIII monitored circuit using the language of Kraus operators and then describe the details of the simulation, including the evolution time, ensemble size, the domain-wall dynamics, and the braiding protocol.

\subsection{Kraus operator description}
In the class-DIII monitored circuit as shown in Fig.~\ref{fig:DIII}(a), each gate, including all the effects of the probabilistic application of the unitary gates and the measurement, can be captured by an
ensemble of Kraus operators as $\mathcal{M}^{\text{DIII}}(p)=\left\{ K_{+}^{\text{DIII}}(p), K_{-}^{\text{DIII}}(p),K_{U}^{\text{DIII}}(p,\theta) \right\}$, where
\begin{equation}\label{eq:kraus_DIII}
    \begin{split}
        K_{\pm}^{\text{DIII}}(p)&=\sqrt{p}\frac{1\pm \ii \hat{\gamma}_i \hat{\gamma}_{i+1} }{2},\\
        K_{U}^{\text{DIII}}(p,\theta)&=\sqrt{1-p} \exp(\theta \hat{\gamma}_i \hat{\gamma}_{i+1} ).
        \end{split}
\end{equation}
Here, $K_{\pm}^{\text{DIII}}$ captures the wave function collapse associated with the two measurement outcomes $\ii \hat{\gamma}_i \hat{\gamma}_{i+1}=\pm1$ while $K_{U}^{\text{DIII}}$ describes the unitary evolution. As will be clear below, their prefactors $\sqrt{p}$ and $\sqrt{1-p}$ encode the probabilities of applying a measurement and a unitary operation at the given quantum gate.

Given a (normalized) state $|\psi\rangle$ undergoing this quantum gate action, each $K \in\mathcal{M}^{\text{DIII}}$ corresponds to a quantum trajectory in which the state evolves as 
\begin{align}
    |\psi\rangle \rightarrow \frac{K\ket{\psi}}{\norm{\ket{K\psi} }}.
\end{align}
The Born-rule probability for this quantum trajectory is given by $\langle \psi|K^\dag K |\psi\rangle$. One can see that prefactors $\sqrt{p}$ and $\sqrt{1-p}$ in Eq. \eqref{eq:kraus_DIII} enter $\langle \psi|K^\dag K |\psi\rangle$ as the probabilities for applying a measurement or a unitary operation. This Kraus operator ensemble $\mathcal{M}^{\text{DIII}}$ satisfies the positive-operator valued measure (POVM) condition, i.e., $\sum_{K\in\mathcal{M}^{\text{DIII}}}K^\dagger K=\mathds{1}$, which guarantees the normalization of probability.

\subsection{Evolution time and ensemble size}
In Fig.~\ref{fig:DIII}(b), we show the average steady-state MI $\overline{I_{A,B}}$.
Here, the steady state is reached by simulating the circuit following the Born rule up to $L$ time steps.
The average MI $\overline{I_{A,B}}$ is obtained by first averaging over the location of the antipodal region $A$ and $B$ while keeping the size of the two regions and their relative distance fixed, and then sampling across 1000 realizations of different circuits and measurement outcomes.

Both evolution time and ensemble size also apply to the class-A and class-AIII monitored circuit, as shown in Fig.~\ref{fig:A} and Fig.~\ref{fig:AIII}, respectively.

\subsection{Domain-wall dynamics}\label{sec:DW}
In Fig.~\ref{fig:DIII_DW}, we construct DWs in spacetime to explore area-law phases with different topologies and the DTDMs in between.
This DW is obtained by different parametrizations of the ensemble of Kraus operators. 
For example, in Fig.~\ref{fig:DIII_DW}(a) in the main text, the two class-DIII area-law phases are obtained by $p\equiv p_{\text{odd}}=0.1$ and $0.9$, respectively.
Therefore, we can parametrize a spacetime-dependent $p(i,t)$ to interpolate these two $p$ values.
In addition, to avoid DWs from changing abruptly, we smoothen the change across the two regions with different parametrizations using a $\tanh(k i)$ function. 
We choose $k=0.5$ in numerics.
This same logic also applies to the class-A monitored dynamics, as shown in Fig.~\ref{fig:A}(c)

\subsection{Braiding protocol}\label{sec:braiding}
In Fig.~\ref{fig:braiding}, we utilize a T-junction model consisting of three $L$-site Majorana chains in open boundary conditions (assuming $L$ to be even) to swap DTDMs $\alpha_2$ and $\beta_1$. 
Below, we show the details of the braiding protocol, which consists of the following four stages, and then explain the pairwise MI in each stage. 
\begin{enumerate}
    \item We initialize the T-junction into a product state on all three chains, where each chain has a fixed parity on odd Majorana pairs $\left( \hat{\gamma}_{2i-1}^{\mathfrak{a}},\hat{\gamma}_{2i}^{\mathfrak{a}} \right)$ with $\mathfrak{a}=\left\{ \mathfrak{A},\mathfrak{B},\mathfrak{C} \right\}$. 
    At time $t=t_0$, we create two DWs by configuring $p(i,t)$ to dimerize even pairs of Majorana modes, $\left( \hat{\gamma}_{2i}^{\mathfrak{a}},\hat{\gamma}_{2i+1}^{\mathfrak{a}} \right)$, within the region enclosed by the two DWs, in chain $\mathfrak{a}=\left\{ \mathfrak{A},\mathfrak{B}\right\}$. 
    Without loss of generality, we place the DWs at $\frac{L}{4}$  and $\frac{3L}{4}$, creating two DTDMs, corresponding to unmeasured Majorana zero modes, $\alpha_1$, $\alpha_2$ ($\beta_2$, $\beta_1$) in chain $\mathfrak{A}$ ($\mathfrak{B}$). Each pair of DTDMs is maximally entangled within the chain, leading to all three pairwise MIs being 0. (The site index can be chosen to run from the outermost to the center of the T-junction.)
    \item At time $t=t_1$, we move the DTDM $\alpha_2$ from chain $\mathfrak{A}$ to chain $\mathfrak{C}$ by moving the DW from chain $\mathfrak{A}$ to chain $\mathfrak{C}$, treating the two chains as a single chain.
    Therefore, we need to measure the two Majorana sites at the intersection between chains $\mathfrak{A}$ and $\mathfrak{C}$, $\hat{\gamma}_{L}^\mathfrak{A}$ and $\hat{\gamma}_{L}^\mathfrak{C}$, leading to the MI $I_{\mathfrak{A},\mathfrak{C}}$ being $2\log2$ at $t_1$, as the long-range entangled Majorana pair contributes $\log2$ and the local entangled pair at the intersection between $\mathfrak{A}$ and $\mathfrak{C}$ contributes another $\log2$.
    \item At time $t=t_2$, we move the DTDM $\beta_1$ from chain $\mathfrak{B}$ to chain $\mathfrak{A}$ using the same procedure as in the previous step. 
    To make $\beta_1$ leave chain $\mathfrak{B}$ and land on chain $\mathfrak{A}$, we need to measure the innermost Majorana modes between chains $\mathfrak{B}$ and $\mathfrak{C}$, leading to the MI $I_{\mathfrak{B},\mathfrak{C}}$ being $\log2$ coming from the local entangled pair at the intersection, and both $I_{\mathfrak{A},\mathfrak{B}}$ and $I_{\mathfrak{A},\mathfrak{C}}$ being $\log2$ from their respective long-range entangled DTDMs. 
    \item At time $t=t_3$, we move the DTDM $\alpha_2$ from chain $\mathfrak{C}$ to chain $\mathfrak{B}$ using the same procedure by treating the two chains as a single chain. Since chain $\mathfrak{C}$ returns to its initial state, the only entangling chains are $\mathfrak{A}$ and $\mathfrak{B}$ with the MI of $2\log2$.
\end{enumerate}
For the convenience of the description above, we assume an ``ideal'' case in a measurement-only limit.
However, this assumption is not essential--- In Fig.~\ref{fig:braiding}, we use a more generic configuration by programming $p(i,t)$ to vary between 0.15 and 0.85 with the smoothness of the DW $k$ being $0.5$, to demonstrate that the braiding protocol is robust in the presence of symmetry-allowed unitary operators.

\section{Class-A monitored circuit}
In this section, we extend the scope beyond the class-DIII monitored circuit to include a U(1) symmetry. 
By embedding each complex fermion into a pair of Majorana fermions, the circuit can be efficiently simulated within a unified covariance matrix formalism. As discussed in the main text, the class-A monitored circuit requires both a U(1) symmetry and an anti-unitary chiral symmetry ${\cal C}$. We take a two-step process in the following. We will first discuss the monitored dynamics with only the U(1) symmetry required, which belongs to symmetry class AIII. The ensemble of Kraus operators in this class will be provided. To obtain the monitored dynamics in symmetry class A, we post-select the quantum trajectories in class AIII that respect an extra anti-unitary chiral symmetry ${\cal C}$. As it turns out, post selections are inevitable for class-A free-fermion monitored circuits. For class A, we present the Kraus operator description for the onsite unitary gate and the measurement gate as shown in Fig.~\ref{fig:A}(a), and then generalize the onsite unitary gate to a more generic long-range unitary gate.
Finally, we show that the robustness of the DTDMs in the class-A monitored circuit is {compromised} when the required symmetries are broken.

\subsection{Class-AIII monitored circuit}\label{sec:AIII}
\begin{figure}[htbp]
    \centering
    \includegraphics[width=3.4in]{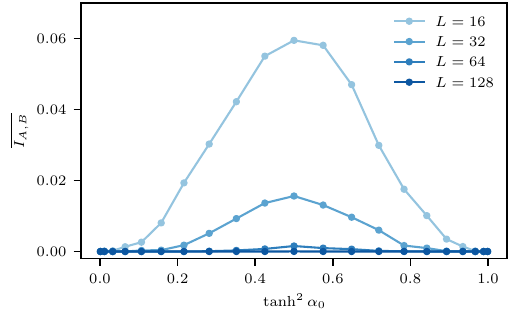}
    \caption{
        Average steady-state mutual information $\overline{I_{A,B}}$ as a function of the measurement strength $\alpha_0$ in class-AIII monitored circuit, showing only one area-law phase.
    }
    \label{fig:AIII}
\end{figure}

We start with a class-AIII monitored circuit, which can be obtained by imposing a U(1) symmetry to the class-DIII monitored circuit.
The transfer matrix $\mathfrak{t}_{\text{AIII}}$ for a 2D class AIII ensemble is in the complexified general linear group $\text{GL}(N,\mathbb{C})$~\cite{ludwig2013lyapunov}.
Therefore, we can parametrize it as $\mathfrak{t_{\text{AIII}}}=\exp(\mu_\text{AIII})$, with $\mu_\text{AIII} \in\mathbb{C}^{N\times N}$. 
In order to later connect to class-A monitored circuit, we can choose $N=2$ and construct the ensemble of Kraus operators as $\mathcal{M}^{\text{AIII}}(\alpha,\theta_1,\theta_2,i,j)=\{ K_{s_+,s_-}^{\text{AIII}}(\alpha,\theta_1,\theta_2,i,j) | s_\pm \in \left\{ -1,+1 \right\}\}$, where 
\begin{equation}\label{eq:kraus_AIII}
    \begin{split}
        &K_{s_+,s_-}^{\text{AIII}}(\alpha,\theta_1,\theta_2,i,j)
        =\frac{1}{2\cosh\alpha}e^{s_+\alpha \left( c_+^\dag c_+ -\frac{1}{2} \right)+s_-\alpha \left( c_-^\dag c_- -\frac{1}{2} \right)} \\
        &e^{\ii\theta_1\left( c_{i,\sfA}^\dagger c_{i,\sfA}-\frac{1}{2} \right)}
        e^{\ii\theta_2\left( c_{j,\sfB}^\dagger c_{j,\sfB}-\frac{1}{2} \right)}.
    \end{split}
\end{equation}
Here, the first line in Eq.~\eqref{eq:kraus_AIII} corresponds to a weak measurement with strength $\alpha\in[0,\infty)$ on the fermion modes $c_{\pm}^\dagger =\frac{1}{\sqrt{2}}\left( c_{i,\sfA}^\dagger \pm c_{j,\sfB}^\dagger\right)$ for $s_+=-s_-=\pm1$, or on the fully-occupied (empty) fermion modes for $s_+=s_-=+1$ ($s_+=s_-=-1$).
The second line in Eq.~\eqref{eq:kraus_AIII} corresponds to unitary gates where $\theta_1, \theta_2$ are uniformly distributed. 
The prefactor $\left( 2\cosh\alpha \right)^{-1}$ is to ensure the POVM condition.
Due to the absence of topological phases in 2D class AIII, the class-AIII monitored circuit always shows an area-law phase corresponding to a trivial disordered insulator as shown in Fig.~\ref{fig:AIII}. Here, we follow the same staggered pattern of $\alpha_0$ and $\alpha_1$ as in Fig.~\ref{fig:A} but with a different set of Kraus operators defined in Eq.~\eqref{eq:kraus_AIII}.

\subsection{Class-A monitored circuit with onsite unitary gates}\label{sec:A}
In Fig.~\ref{fig:A}, we study the class-A monitored circuit. 
{The prescription is that we start with a class-AIII monitored circuit by imposing U(1) symmetry to the class-DIII circuit. 
On top of this class-AIII circuit, we impose an additional anti-unitary chiral symmetry to we obtain the class A circuit.}
Below, we show that this is equivalent to post-select the outcomes that respect the chiral symmetry in the class-AIII monitored circuit.

We start with the formal description of transfer matrix $\mathfrak{t_{\text{A}}}$ for a class A ensemble, which is an indefinite unitary group $\text{U}(N,N)$~\cite{ludwig2013lyapunov}, 
The minimal example is $\text{U}(1,1)$ with each basis for the two sublattice $\sfA$ and $\sfB$, leading to the constraint as $\mathfrak{t_{\text{A}}}^\dagger \sigma_z \mathfrak{t_{\text{A}}} = \sigma_z$, with $\sigma_z$ being the Pauli Z matrix. One can show that this constraint on the single-particle transfer matrix $\mathfrak{t}_{\text{A}}$ exactly matches the requirement of an anti-unitary chiral symmetry ${\cal C}$, defined by $ c_{i,\sfA} \rightarrow -c_{i,\sfA}^\dagger,  c_{i,\sfB} \rightarrow c_{i,\sfB}^\dagger, \ii \rightarrow -\ii$, as introduced in the main text.

Here, it can be shown that by setting $s_+=-s_-$ in Eq.~\eqref{eq:kraus_AIII}, the aforementioned constraint for the transfer matrix $\mathfrak{t_{\text{A}}}$ can be satisfied, leading to the ensemble of Kraus operators for the class-A monitored as $\mathcal{M}^{\text{A}}(\alpha,\theta_1,\theta_2, i,j)=\{ K_{\pm}^{\text{A}}(\alpha,\theta_1,\theta_2,i,j)\}$
where 
\begin{equation}\label{eq:kraus_A}
    \begin{split}
        &K_{\pm}^{\text{A}}(\alpha,\theta_1,\theta_2,i,j)= \frac{1}{2\cosh\alpha}e^{\pm\alpha\left(c_{i,\sfA}^\dagger c_{j,\sfB} + \text{h.c.} \right)} \\
        & e^{\ii\theta_1\left( c_{i,\sfA}^\dagger c_{i,\sfA}- \frac{1}{2} \right)}e^{\ii\theta_2\left(c_{j,\sfB}^\dagger c_{j,\sfB}- \frac{1}{2}\right)}.
    \end{split}
\end{equation}
This is the ensemble of Kraus operators used in Fig.~\ref{fig:A}.

From Eq.~\eqref{eq:kraus_AIII} for class AIII to Eq.~\eqref{eq:kraus_A} for class A, we exclude the two possible outcomes with $s_+=s_-=\pm 1$, corresponding to the fully-empty and fully-occupied state, respectively.
In fact, even considering a generic U($N$,$N$) for the transfer matrix, which involves measurement of $N$ pairs of complex fermions, it becomes inevitable to sacrifice POVM for the chiral symmetry $\mathcal{C}$ in class-A monitored circuit, because the fully-empty state and fully-occupied state are always decoupled and do not respect the chiral symmetry $\mathcal{C}$ by themselves.

Due to the absence of POVM, we adopt a post-selection Born rule by renormalizing the Born probabilities for the two fermion modes $\frac{1}{\sqrt{2}}\left( c_{i,\sfA}^\dagger \pm c_{j,\sfB}^\dagger  \right)$ as 
\begin{equation}\label{eq:prob_A}
    \begin{split}
        &p\left( K_{\pm}^{\text{A}}(\alpha,\theta_1,\theta_2,i,j)  \right)\\
        &=\frac{\expval{\left( K_{\pm}^{\text{A}}(\alpha,\theta_1,\theta_2,i,j) \right)^\dagger K_{\pm}^{\text{A}}(\alpha,\theta_1,\theta_2,i,j)}}{\sum\limits_{s=\pm 1} \expval{\left( K_{s}^{\text{A}}(\alpha,\theta_1,\theta_2,i,j) \right)^\dagger K_{s}^{\text{A}}(\alpha,\theta_1,\theta_2,i,j)}}.
    \end{split}
\end{equation}

\begin{figure}[htbp]
    \centering
    \includegraphics[width=3.4in]{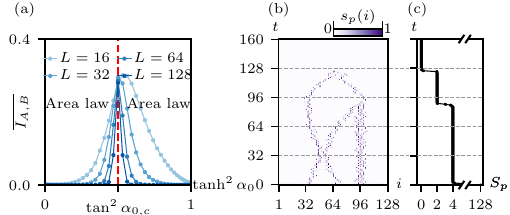}
    \caption{
        (a) Phase diagram of steady-state class-A monitored circuit with longer-range unitaries in Eq.~\eqref{eq:kraus_A_nonlocal}. Average steady-state mutual information $\overline{I_{A,B}}$ as a function of the measurement strength $\alpha_0$,  with $\tanh^2{\alpha_0}+\tanh^2\alpha_1=1$, showing two topologically different area-law phases separated by a critical point.
        (b-c) Entanglement contour and total entanglement entropy (in binary logarithm) of the physical chain in a \textit{typical} quantum trajectory. The peaks in the entanglement contour indicate the spacetime position of the dynamical topological domain-wall modes.
        }
    \label{fig:A_nonlocal}
\end{figure}
\subsection{Class-A monitored circuit with longer-range unitary gates}\label{sec:longrange}
In Fig.~\ref{fig:A}, we choose the onsite unitary gate in the Kraus operators defined in Eq.~\eqref{eq:kraus_A}. 
However, this choice is merely a convenient choice for illustrative purposes. 
In principle, any generic (nonlocal) unitary gate acting on a range-$R$ dimer from the same sublattice respects the chiral symmetry and thus should demonstrate the same classification of area-law phases as in Fig.~\ref{fig:A}(b).
Therefore, the Kraus operators can be generalized with the longer-range unitary gate as per
\begin{equation}\label{eq:kraus_A_nonlocal}
    \begin{split}
        &\tilde{K}_{\pm}^{\text{A}}(\alpha,\theta_1,\theta_2,i,j)= \frac{1}{2\cosh\alpha}e^{\pm\alpha\left( c_{i,\sfA}^\dagger c_{j,\sfB} + \text{h.c.} \right)}  \\
        &e^{\ii\frac{\theta_1}{2}\left( c_{i,\sfA}^\dagger c_{j,\sfA} - c_{i,\sfA}c_{j,\sfA}^\dagger \right)}e^{\ii\frac{\theta_2}{2}\left(c_{i,\sfB}^\dagger c_{j,\sfB} - c_{i,\sfB}c_{j,\sfB}^\dagger\right)}.
    \end{split}
\end{equation}

The phase diagram of the class-A monitored circuit using longer-range unitary gates is shown in Fig.~\ref{fig:A_nonlocal}(a).
The difference between Fig.~\ref{fig:A_nonlocal}(a) and Fig.~\ref{fig:A}(b) is merely quantitative, which does not change the existence of two area-law phases with different topologies separated by a critical point. 
However, due to the long-range nature of the unitary gate in Eq.~\eqref{eq:kraus_A_nonlocal}, it is harder to reach the IR limit, and thus we reduce the variance of the longer-range unitary gate from $\theta_{1,2}\in[-\pi,\pi]$ to $[-\frac{\pi}{4},\frac{\pi}{4}]$, which effectively decreases the correlation length~\cite{jian2020measurementinduced}.

Following the same procedure as the DW dynamics in the class-A monitored circuit in Figs.~\ref{fig:A}(c-e), we can perform the same DW dynamics with the longer-range unitary gate in Eq.~\eqref{eq:kraus_A_nonlocal} as shown in Figs.~\ref{fig:A_nonlocal}(b-c). 
We use the same spacetime profile of $r(i,t)$ as in Fig.~\ref{fig:A}(c), and verify the same DTDMs with $\mathbb{Z}$ classification, consistent with the onsite unitary gate in the main text.

\begin{figure}[htbp]
    \centering
    \includegraphics[width=3.4in]{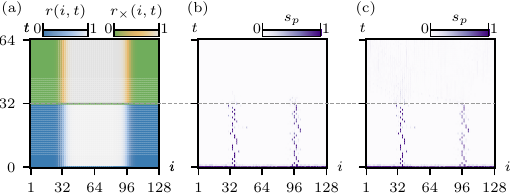}
    \caption{
        (a) Parametrization of chiral-symmetry-preserving $r(i,t)$ before time $t=32$, and chiral-symmetry-breaking $r_{\times}(i,t)$ after time $t=32$
        (b-c) Entanglement contours (in binary logarithm) in the class-A monitored circuit for $t<32$. 
        For $t>32$, the chiral symmetry is broken by (b) measurements on the same sublattice or (c) unitary gates on different sublattices, both leading to the compromising of the DTDMs.
        Ensembles of Kraus operators are (b) Eq.~\eqref{eq:kraus_A} and (c) Eq.~\eqref{eq:kraus_no_chiral}.
        }
    \label{fig:A_breakC}
\end{figure}
\subsection{Compromised domain-wall modes under broken symmetries}\label{sec:symm_break}
In previous sections for class-A monitored circuits, we ensure that unitary gates (measurements) always act on the pair of complex fermion sites within the same sublattice (across the opposite sublattices) to respect the chiral symmetry $\mathcal{C}$.
In this section, we will show that if we include the measurements within the same sublattice, or the unitary gates across the sublattices, the DTDMs will be comprised.
This directly manifests that the DTDMs in Fig.~\ref{fig:A} are indeed protected by the symmetry from being quenched by the local symmetry-allowed disorder.

\subsubsection{Measurements on the same sublattice}
We start with the class-A monitored circuit, and design the spacetime profile of $r(i,t)$ to create two DWs at $i=\frac{L}{4}$ and $i=\frac{3L}{4}$ initially for time $t<32$ as shown in Fig.~\ref{fig:A_breakC}(a).
After $t=32$, we introduce the Kraus operators, which allow measurement on complex fermion sites within the same sublattice.
Therefore, we can choose Eq.~\eqref{eq:kraus_A} as the ensemble of Kraus operators, with $r_\times \equiv i-j$ (the subscript $\times$ is to emphasize that the Kraus operators break the chiral symmetry $\mathcal{C}$).
Here, we still parametrize $r_{\times}(i,t)$ to have two ``DWs'' at $i=\frac{L}{4}$ and $i=\frac{3L}{4}$ for a direct comparison with the $r(i,t)$, though strictly speaking, both regions are now in the same type of area-law phases.

We present the EC to track the DTDMs in Fig.~\ref{fig:A_breakC}(b). We find that DTDMs are immediately comprised after introducing the measurements on the same sublattice.

\subsubsection{Unitary gates on different sublattices}
Apart from measurements on the same sublattice, we can also introduce the unitary gates that can couple the different sublattices, breaking the chiral symmetry $\mathcal{C}$ as well. 
These chiral-symmetry-broken Kraus operators can be chosen as 
\begin{equation}\label{eq:kraus_no_chiral}
    \begin{split}
        &K_{\pm,\times}^{}(\alpha,\theta_1,\theta_2,i,j)= \frac{1}{2\cosh\alpha}e^{\pm\alpha\left( c_{i,\sfA}^\dagger c_{j,\sfB} + \text{h.c.} \right)}  \\
        &e^{\ii\frac{\theta_1}{2}\left( c_{i,\sfA}^\dagger c_{i,\sfB} - c_{i,\sfB}c_{i,\sfA}^\dagger \right)}e^{\ii\frac{\theta_2}{2}\left(c_{j,\sfA}^\dagger c_{j,\sfB} - c_{j,\sfB}c_{j,\sfA}^\dagger\right)},
    \end{split}
\end{equation}
with $r_\times\equiv i-j$ defined the same as before.

In Fig.~\ref{fig:A_breakC}(c), we present the EC for such chiral-symmetry-breaking Kraus operators, and find that the DTDMs are also compromised once unitary gates on different sublattices are applied after $t>32$. 

\end{document}